\documentclass[aps,pre,twocolumn,amsmath,amssymb,amsfonts]{revtex4-1}
\usepackage{epsfig}
\usepackage{graphicx}
\usepackage{dcolumn}
\usepackage{bm}\usepackage[titletoc]{appendix}

\usepackage{graphicx}
\usepackage{subfigure}
\usepackage{amsfonts}
\usepackage{amsgen}
\usepackage{amsbsy}
\usepackage{epstopdf}
\usepackage{caption}
\usepackage{tabularx}
\usepackage{tabu}
\usepackage{xcolor}
\usepackage{nicefrac}
\usepackage{enumitem}
\usepackage[normalem]{ulem} 

\newcommand{\e}[1]{\ensuremath{\times 10^{#1}}}

\begin{document}

\author{Itzhak Fouxon$^{1,2}$} \email{itzhak8@gmail.com}  \author{Zhouyang Ge$^3$}\email{zhoge@mech.kth.se} \author{Luca Brandt$^3$}\email{luca@mech.kth.se} \author{Alexander Leshansky$^1$}\email{lisha@tx.technion.ac.il}
\affiliation{$^1$ Department of Chemical Engineering, Technion, Haifa 32000, Israel}
\affiliation{$^2$ Department of Computational Science and Engineering, Yonsei University, Seoul 120-749, South Korea}
\affiliation{$^3$ Linn\'{e} FLOW Centre and SeRC (Swedish e-Science Research Centre), KTH Mechanics, SE-100 44 Stockholm, Sweden}

\begin{abstract}

We construct a boundary integral representation for the low-Reynolds-number flow in a channel in the presence of freely-suspended particles (or droplets) of arbitrary size and shape. We demonstrate that lubrication theory holds away from the particles at horizontal distances exceeding the channel height and derive a multipole expansion of the flow which is dipolar to the leading approximation. We show that the dipole moment of an arbitrary particle is a weighted integral of the stress and the flow at the particle surface, which can be determined numerically. We introduce the equation of motion that describes hydrodynamic interactions between arbitrary, possibly different, distant particles, with interactions determined by the product of the mobility matrix and the dipole moment. Further, the problem of three identical interacting spheres initially aligned in the streamwise direction is considered and the experimentally observed ``pair exchange" phenomenon is derived analytically and confirmed numerically. For non-aligned particles, we demonstrate the formation of a configuration with one particle separating from a stable pair. Our results suggest that in a dilute initially homogenous particulate suspension flowing in a channel the particles will eventually separate into singlets and pairs.

\end{abstract}

\title{Integral representation of channel flow with interacting particles}
\maketitle

\section{Introduction}

Hydrodynamic interactions among particles flowing in the fluid confined between two parallel walls at low Reynolds number have recently attracted a considerable attention \cite{carba,d1,d2,d3,d4,d5,d6,d7,anom,7,tlusty,tl2006,tl2012,is,sb,f1,f2,f3,f4,f5,f7,tl2014,tabeling,tab0}. The case of particles driven by thermal noise in the absence of a macroscopic flow was studied in \cite{carba,d1,d2,d3,d4,d5,d6,d7,anom,7,tlusty}. The hydrodynamic interactions cause long-range correlations
in their diffusive motions
that are measurable even at distances ten times larger than the particle size \cite{carba,anom}. In the case of
pressure-driven Poiseuille or shear flow the particles are, in addition, dragged by the flow \cite{f1,f2,f3,f4,f5,f7,tl2006,tl2012,tl2014,tabeling,tab0,is}.

Identical particles at similar positions inside the channel move at the same velocity if not for hydrodynamic interactions. These interactions induce particle relative motions, which can result in considerable changes of their configuration inside the channel.
In the case of a large number of particles the interactions cause also chaotic collisions among the particles \cite{tl2014}.

Theoretical progress has mainly relied on the observation that the far flow caused by a particle confined in a channel is a dipolar flow decaying quadratically with the distance \cite{anom}. For disk-like particles with thickness close to the channel height $h$, the dipolar flow and its moments were derived from lubrication theory in \cite{tlusty}. The dipolar flow holds at distances much larger than the disk radius, where it gives also the leading order hydrodynamic interactions among particles \cite{tlusty,tl2006,tl2012,tl2014}.

It was observed in \cite{tl2012}, however, that hydrodynamic interactions of pancake-like disks can also be described at much smaller distances between the disks where dipolar approximation breaks down, yet lubrication theory still holds \cite{Batchelor,lubr,szeri,bruce}. This theory predicts that at distances from the particle boundary much larger than $h$ the depth-averaged flow is an ideal two dimensional flow with potential obeying the Laplace equation. The boundary condition (b.\ c.), derived somewhat heuristically, is the usual ideal flow b.\ c.\ prescribing  the velocity component normal to the particle surfaces \cite{ll}, which allowed to find the hydrodynamic interactions of two close disks, see \cite{tl2012}. Moreover, it was observed that the non-rigidity of the particles makes the lubrication theory valid up to distances from the particles smaller than $h$. The calculation of the hydrodynamic interactions for disks of different radii requires solving the Laplace equation with the help of bipolar coordinates, see \cite{is}.

Recently, a practical application of hydrodynamic interactions among particles in a channel has been proposed. In particular, it is suggested that the combined action of adhesive (non-hydrodynamic) forces and hydrodynamic interactions between microdroplets can result in the formation of regular particle clusters and can thus be potentially used for the production of new materials \cite{tab0,tabeling}. The hydrodynamic forces are believed to be a significant factor in these structure formation. Though the particles
forming the structure are in a close proximity in the experiments mentioned above, the hydrodynamic interactions are described phenomenologically by a dipolar flow, formally only valid at larger distances. Despite the use of the far-field dipolar flow beyond its domain of validity, the numerical simulations  in \cite{tab0} showed very good agreement with the experimental results \cite{tabeling}. The above motivates the need for the detailed theoretical study of hydrodynamic interactions among particles in narrow channels.

In this work, we introduce a boundary integral representation of the channel flow in the presence of freely suspended particles. The particles can be rigid or soft (droplets).
The representation does not depend on the particle equation of motion, defined by inertia.
Boundary integral representations are known to be useful in unconfined flows and can also be applied to confined geometries  \cite{hb,ps}.
The flow is here expressed as the sum of the undisturbed Poiseuille flow and an integral over the surfaces of all particles, where the particles can have arbitrary shapes.
The derivation is performed for a pressure-driven flow, but identical considerations can be applied to shear flows.

Our representation results in a formula for calculation of the dipole moment, which was previously available only for the case of disk-like particles. The moment is given in terms of a weighted integral of the
stress tensor and the  flow over the surface(s) of the particle(s). Once this integral is numerically tabulated, the result can be used to approximate the flow in different configurations. Here, we perform simulations for the case of neutrally buoyant rigid spherical particles and compute the integral for different positions of the particle center and different ratios of the particle radius to the channel height, i.e.\ different confinements.

We use this new integral representation to show that the lubrication theory holds at the particle near proximity, closer than what typically expected. As an example, we solve the problem of three aligned particles moving along the line defined by their centers and  the case of three nonaligned particles.
We conjecture that this solution is the attractor to which the long-time evolution of arbitrary initial condition converges. We conclude by proposing a mean field description of strong hydrodynamic interactions of close
particles in a dense suspension.

\section{Integral representation for channel flow with particles}

\begin{figure}[t!]
 \begin{center}
\includegraphics[width=.9\columnwidth]{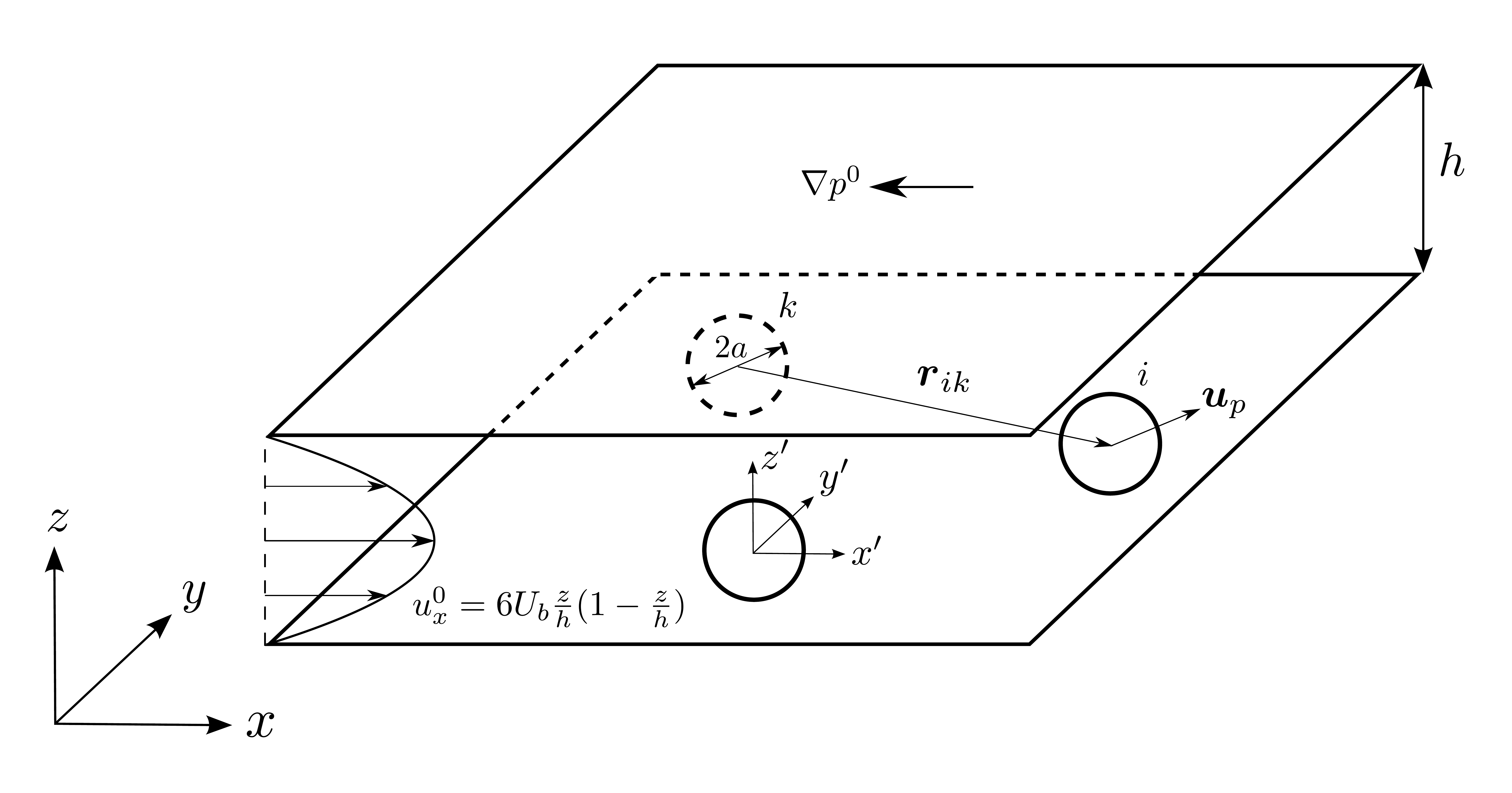}
\end{center}
  \caption{Schematic configuration of spherical particles flowing in the pressure-driven (Poiseuille) flow in a channel.}
 \label{fig: setup}
\end{figure}

In this Section we derive the boundary integral representation for channel flow in the presence of an arbitrary number of particles of arbitrary shape (see Fig.\ \ref{fig: setup} where spherical particles are shown for illustration). It is assumed that the Reynolds number is low and the Stokes equations hold. The
derivation uses the reciprocal theorem with the reciprocal flow given by the Stokeslet in a channel \cite{LironMochon}, similarly to the derivations in infinite space, see e. g. \cite{ps}.
In this Section we make no assumptions on the form of the equation of motion of the particles which may change according to the relevance of inertia. The particles can be rigid, droplets or, e.g.\ viscoelastic.

The undisturbed flow $\bm u^0$, in the absence of  particles, is the Poiseuille flow driven by the constant pressure gradient $\nabla p^0$,
\begin{eqnarray}&&
u^0_x=\frac{z(z-h)\nabla_x p^0}{2\eta},\ \ \nabla p^0=\eta\nabla^2\bm u^0.\label{ps}
\end{eqnarray}
where $\eta$ is the fluid viscosity, $z$ is the vertical coordinate and $h$ is the channel height. The flow is in $x-$direction, $\nabla_x p^0=-|\nabla_x p^0| \bm{\hat x}$. In the presence of a freely suspended particle we look for the solution of,
\begin{eqnarray}&&
\nabla p=\eta\nabla^2\bm u,\ \ \nabla\cdot\bm u=0,\ \ \ \ \bm u(z=0)=\bm u(z=h)=0,\nonumber\\&&
u_x(\infty)=\frac{z(z-h)\nabla_x p^0}{2\eta},
\label{ds}
\end{eqnarray}
which holds outside the particle. The flow is completely determined when solving for the particle motion, i.e.\ knowing the instantaneous particle position as it determines the boundary condition $\bm u_{S}(\bm x)$ on the
 particle surface $S$. In the case of a rigid particle, $\bm u_{S}(\bm x)=\bm v+\bm \omega\times (\bm x-\bm y)$ where $\bm v$ and $
\bm \omega$ are the particle translational and angular velocities  and $\bm y=(x_p, y_p, z_p)$ is the coordinate of the particle center of mass. The velocities $\bm v$ and $\bm \omega$ are determined by the solution of the equation for the particle motion coupled with the flow. These velocities could be time-independent as in the case of the steady motion of a neutrally buoyant rigid particle or the case of a non-neutrally buoyant particle after sedimentation when reaching the bottom wall (the theoretical determination of these velocities is impossible generally because of the interaction with the walls). These velocities can also be time-dependent as in the case of
a transient flow  or the gravitational settling of a non-neutrally buoyant particle. If several particles are considered, a time-dependent configuration can be induced by their interactions. We assume here that the time variations are not fast so that the unsteady time-derivative term of the Navier-Stokes equations is negligible (for the steady motion of one particle the time-derivative is the spatial derivative of the flow along the streamline which is small  because of the smallness of the Reynolds number). In the case of droplets the boundary condition on the surface is determined by matching with the inner flow. However there is no need for solving for this inner flow since the detailed form of $\bm u_{S}(\bm x)$ is irrelevant for the derivation of the present representation. The generalization of the problem to the case of many particles is obvious.

{\it Implications of lubrication theory.}---Some conclusions on the flow at distances from the particle much larger than the channel height $h$ can be obtained from lubrication theory \cite{Batchelor,lubr,szeri,bruce}. The lubrication theory predicts that at these distances,
\begin{eqnarray}&&
\bm u=\frac{z(z-h)\nabla p}{2\eta},\label{lubr}
\end{eqnarray}
with a certain $z-$independent $p$. Clearly, at large distances, $p\approx p^0$ at the leading order. The depth-averaged velocity $\bm u_d$ is the ideal potential two-dimensional flow,
\begin{eqnarray}&&
\bm u_d=\nabla\phi,\ \ \phi=-\frac{h^2 p}{12\eta},\ \ \nabla^2\phi=0.\label{dlp0}
\end{eqnarray}
In some cases this helps determining the flow completely.

{\it Large disks.}---Large non-wetting droplets squeezed between the walls of a Hele-Shaw cell have pancake-like shapes. These can be modeled as disks with radius $a\gg h$ and height close to $h$, i.e., they almost fill the entire channel height \cite{tlusty,tl2006,tl2012,tl2014}. In this case, one can use Eqs.~(\ref{lubr})-(\ref{dlp0}) at distances from the body that are much larger than $h$ but much smaller than $a$. Thus the ideal flow holds outside the narrow boundary layer near the particle surface whose characteristic size $l_0$ is of the order $h$, much smaller than the particle horizontal size $a$. We call the layer containing the flow vorticity the viscous layer and assume that there is no flux of mass through the surface of the droplet, which keeps its shape and volume. Thus, in the frame of reference moving with the disk, the normal
velocity component is zero at the outer boundary of the layer and the ideal flow outside the layer is determined uniquely by this boundary condition. However the geometry of the layer is not always known and the complete determination of the flow not possible. If we are only interested in the flow outside the immediate $l_0-$vicinity of the viscous layer,  we can set the boundary condition for the ideal flow on the disk itself, exploiting the fact that $l_0\ll a$. The flow is then found as the dipole potential \cite{tlusty},
\begin{eqnarray}&&
\phi=-\frac{h^2 p^0}{12\eta}-\frac{\bm d\cdot{\hat r}}{r} ,\ \ \nabla^2\phi=0,\ \ \bm d=a^2\bm v,\label{dl}
\end{eqnarray}
where $\bm d$ is the dipole moment, $\bm v=-v\bm{\hat x}$ is the difference between the disk velocity and $-h^2\nabla p_0/(12\eta)$. It is readily seen that the normal, radial component of the velocity $\nabla \phi$ on the surface of the disk is $[\bm v-h^2\nabla p^0/(12\eta)]\cdot{\hat r}$. Note that the velocity $v>0$ since the particle moves slower than the fluid.

The tangential velocity component on the outer boundary of the viscous layer obtained from Eq.~(\ref{dl}) does not match the tangential velocity of the droplet surface. In contrast with the normal component, which can be considered almost constant through the viscous layer, the tangential component changes quickly through this layer to match the inner flow at the droplet surface. As example, in the limit of high droplet viscosity, the condition on the disk surface is that the flow is the appropriate superposition of translation and solid body rotation. Similar viscous layers occur for rigid bodies oscillating in the fluid \cite{ll}. The ideal flow was obtained in \cite{tlusty} and here we describe how this ideal flow fits the complete equations for the viscous flow.

{\it Boxes.}---Another case where the flow can be fixed without detailed calculations is the case of a box whose smallest dimension is close to $h$ and the longer dimensions are much larger than $h$. If the box is located in the channel so that the flow is
perpendicular to its longer axis with length $l\gg h$ then far from the ends of the box we find the ideal two-dimensional flow with constant velocity on the line.
The solution for the ideal flow with constant velocity on an infinite linear boundary is the uniform flow. We thus conclude that in the frame moving with the box there is a region of stagnant flow behind the box, whose size is of order $l$.

Generally, the flow can be inferred from the lubrication theory in quite a detail for particles whose horizontal dimensions are much larger than $h$ and whose vertical dimension is close to $h$. The flow outside the narrow viscous layer near the particle surface is $z(z-h)$ times the two-dimensional ideal flow determined by the boundary condition of zero normal velocity on the particle surface. The tangential velocity changes fast across the viscous layer. However, if the particle horizontal dimensions are not large or the dimensions are large but the vertical dimension is not close to $h$, a different approach is needed.

{\it Boundary integral representation from the reciprocal theorem.}---We use the reciprocal theorem \cite{hb,ps} using as the reciprocal flow the solution of \cite{LironMochon} for the point-force or Stokeslet between two parallel plates, i.e.\
\begin{eqnarray}&&
-\nabla p^S+\eta \nabla^2 \bm u^S+\bm g\delta(\bm x-\bm x_0)=0,\ \ \nabla\cdot\bm u^S=0, \label{stokes}\\&& \bm u^S(z=0)=\bm u^S(z=h)=0,\ \ \bm u^S(x^2+y^2\to\infty)=0.\nonumber \end{eqnarray}
Analogously to the flow due to a point-force acting on a viscous fluid in infinite space (e.g. \cite{ps}), the solution depends linearly on the source forcing $\bm g$,
\begin{eqnarray}&&\!\!\!\!\!\!\!\!\!\!
\bm u^S(\bm x)=\frac{1}{8\pi \eta}S_{ik}(\bm x, \bm x_0)g_k, \label{vl}
\end{eqnarray}
where we introduced the tensor $S_{ik}$ independent of $\bm g$. Similarly for the stress tensor of the Stokeslet solution we can write,
\begin{eqnarray}&&
\sigma^S_{ik}(\bm x)\!=-p^S\delta_{ik}+\eta\left(\nabla_k u^S_i+\nabla_i u^S_k\right)\!=\!\frac{T_{ilk}(\bm x, \bm x_0)g_l}{8\pi},\label{st}\\&& \nabla_kT_{ilk}\!=\!-8\pi\delta_{il}\delta(\bm x\!-\!\bm x_0),\ \ T_{ilk}\!\!=\!\!-p_l \delta_{ik}\!+\!\nabla_k S_{il}\!+\!\nabla_i S_{kl},\nonumber
\end{eqnarray}
where $T_{ilk}$ is a third-rank tensor independent of $\bm g$ and we defined the $\bm g-$independent $p_l$ by $p^S=\bm p\cdot\bm g/(8\pi)$.
We use the Lorentz identity,
\begin{eqnarray}&&
\nabla_k\left[u_{i}\sigma^S_{ik}-u^S_i\sigma_{ik}\right]+\bm u\cdot \bm g\delta(\bm x-\bm x_0)=0,\label{lo}
\end{eqnarray}
readily inferred from the Stokes equations with $\sigma_{ik}$ the stress tensor of the flow defined by Eqs.~(\ref{ds}),
\begin{eqnarray}&&
\sigma_{ik}=-p\delta_{ik}+\eta\left(\nabla_iu_k+\nabla_ku_i\right).
\end{eqnarray}
Substituting $\bm u^S$ and $\bm \sigma^S$ from Eqs.~(\ref{vl})-(\ref{st}) in Eq.~(\ref{lo}) and using the above identity we find that,
\begin{eqnarray}&&
8\pi \eta u_{l}(\bm x')\delta(\bm x'-\bm x)=\frac{\partial}{\partial x_k'}\left[S_{il}(\bm x', \bm x)\sigma_{ik}(\bm x')
\right.\nonumber\\&&\left.-\eta u_{i}(\bm x')T_{ilk}(\bm x', \bm x)\right].\label{fd}
\end{eqnarray}
Integrating this equation over $\bm x'$ outside the particles,
\begin{eqnarray}&&\!\!\!\!\!\!\!\!\!\!\!\!\!\!\!\!
u_l\!=\!f_l\!-\!\sum_n\int_{S_n}\!\!\!\frac{S_{il}(\bm x', \bm x)\sigma'_{ik}dS_k'}{8\pi \eta}\nonumber\\&&\!\!\!\!\!\!\!\!\!\!\!\!\!\!\!\!
+\sum_n\int_{S_n}\frac{u_{i}(\bm x')T_{ilk}(\bm x'\!, \!\bm x)dS'_k}{8\pi},\label{in}
\end{eqnarray}
where $\bm f$ is the integral over the far surface at infinity,  $S_n$ is the surface of the $n-$th particle and $dS_k$ is aligned with the outward normal to the particle surface. There is no contribution from the channel boundaries $z=0$ and $z=h$ since both flows vanish there.

We assume that the particles are confined in a finite region so that the flow far from the particles is the Poiseuille flow given by Eq.~(\ref{ps}), see Eq.~(\ref{ds}). Since the Stokeslet decays far from the source, the second and the third terms on the RHS of Eq.~(\ref{in}) decay to zero at large distances from the particles.
 Thus, the asymptotic approach of $\bm u$ to the Poiseuille flow at large distances implies the asymptotic equality of $\bm f$ to the Poiseuille flow. It is readily seen using the asymptotic form of the Stokeslet at large horizontal distances, provided in the next Section, and the asymptotic, Poiseuille, form of the flow, that $\bm f$ is determined by terms independent of the particles, that is terms that would be the same for the case of no particles.
Thus necessarily $\bm f$ is the Poiseuille flow given by Eq.~(\ref{ps}). This is confirmed by the direct calculation in Appendix \ref{f}. We therefore conclude that,
\begin{eqnarray}&&\!\!\!\!\!\!\!\!\!\!\!\!\!\!\!\!
u_l\!=\!\frac{\delta_{lx}z(z\!-\!h)\nabla_x p^0}{2\eta}\!-\!\sum_n\int_{S_n}\!\!\!\frac{S_{il}(\bm x', \bm x)\sigma'_{ik}dS_k'}{8\pi \eta}\nonumber\\&&\!\!\!\!\!\!\!\!\!\!\!\!\!\!\!\!
+\sum_n\int_{S_n}\frac{u_{i}(\bm x')T_{ilk}(\bm x'\!, \!\bm x)dS'_k}{8\pi}.\label{inrep}
\end{eqnarray}
This integral representation of the flow involves no approximations and holds for particles of arbitrary shape. The flow is determined by the values of $\bm u$ and $\sigma_{ik}$ at the particle surface. When the distance between the particles are much larger than their size the hydrodynamic interactions are negligible and $\bm u$ and $\sigma_{ik}$ are approximately those of an isolated particle. However, the current representation works also when the particles are close to each other so that the hydrodynamic interactions change significantly the values of $\bm u$ and $\sigma_{ik}$ at the particle surface.


{\it Simplification for rigid surface.}---The integral representation above simplifies in the case of rigid surfaces when the flow on the surface of the particles is the superposition of translation and solid-body rotation. This is not only the case of rigid particles described after Eq.~(\ref{ds}), but  often also the case of small bubbles where impurities present in the fluid accumulate at the surface making it effectively rigid. In this case, experiments demonstrate that the behavior of these bubbles is similar to that of rigid particles \cite{hb,levich}. In these and similar cases the flow at the particle surface $u_i(\bm x')$ has the form  $c_i+\epsilon_{irm}x'_m{\tilde c}_r$ where both $\bm c$ and $\bm {\tilde c}$ are independent of $\bm x'$. Hence the last term in Eq.~(\ref{inrep}) drops out because  for $\bm x$ outside the particle interior $V_p$,
\begin{eqnarray}&&
\int_{S_p}dS'_k T_{ilk}(\bm x', \bm x)=\int_{V_p} dV \nabla'_k T_{ilk}=0, \\&& \epsilon_{irm}\int_{S}dS'_k x'_mT_{ilk}(\bm x', \bm x)
=\epsilon_{irm}\int_{V_p} dV \delta_{mk}T_{ilk}=0,
\nonumber
\end{eqnarray}
cf.\ \cite{ps}. Note that we made use of the fact that $\epsilon_{irm}$ is antisymmetric over indices $i$, $m$ whereas $T_{ilm}$ is symmetric with respect to those indices. We conclude that in the case of many particles with rigid surfaces,
\begin{eqnarray}&&\!\!\!\!\!\!\!\!\!\!\!\!\!\!\!\!
u_l\!=\!\frac{\delta_{lx}z(z\!-\!h)\nabla_x p^0}{2\eta}\!-\!\sum_n\int_{S_n}\!\!\!\frac{S_{il}(\bm x', \bm x)\sigma'_{ik}dS_k'}{8\pi \eta},\label{inrepmany}
\end{eqnarray}
The representations derived here are thus useful to descrive  the  flow.

\section{Derivation of lubrication theory}

In this Section we demonstrate that Eqs.~(\ref{inrep}) and (\ref{inrepmany}) imply that the predictions of the lubrication theory hold at horizontal distances from the particles larger than $h$. This is less restrictive than the usual condition of applicability of the lubrication theory for distances much larger than $h$, cf.\ \cite{is}. This property comes from the Stokeslet flow which obeys the lubrication theory at distances larger than $h$. We use the representation
\begin{eqnarray}&&\!\!\!\!\!\!\!\!\!\!
S_{il}(\bm x', \bm x)\!=\!\frac{12 z'(h\!-\!z')z(z\!-\!h)\nabla_i\nabla_l  \ln \rho}{h^3}\!+\!{\tilde S}_{il}(\bm x', \bm x),\label{Stokesl}
\end{eqnarray}
where $\rho$ is the length of $\bm \rho\!=\!(x-x', y-y')$ (thus $\nabla_i\nabla_l  \ln \rho=0$ if one of the indices is $z$). It was observed in \cite{LironMochon}
that ${\tilde S}_{il}$ decays exponentially in $\rho$ with exponent at least $\pi/h$, that is the smallness is at least $\exp(-\pi\rho/h)$.
Thus already at $\rho\approx h$ we can discard the last, non-potential term in Eq.~(\ref{Stokesl}). The resulting approximation to the Stokeslet flow,
\begin{eqnarray}&&\!\!\!\!\!\!\!\!\!\!
S_{il}(\bm x', \bm x)\!\approx\! S^0_{il}(\bm x', \bm x)\!= \!\frac{12 z'(h\!-\!z')z(z\!-\!h)\nabla_i\nabla_l  \ln \rho}{h^3},\label{so}
\end{eqnarray}
is the two-dimensional potential flow times $z'(h-z')$, in agreement with the predictions of the lubrication theory, see Eq.~(\ref{lubr}). Note  that $S^0_{il}(\bm x', \bm x)$ is a symmetric function of $\bm x$ and
$\bm x'$, whose dependence on the horizontal coordinates is via the difference $\bm \rho$ only. We consider the corresponding pressure $p^0$ that approximately solves the corresponding Stokes equation
$\nabla'_i p^0_l(\bm x', \bm x)\!=\!\nabla'^2 S^0_{il}(\bm x', \bm x)$, see the second of Eqs.~(\ref{st}). Here $\nabla'_i$ designates the derivative over $x'_i$ and we do not write the $\delta(\bm x'-\bm x)$ term on the RHS. We thus find ($l\neq z$),
\begin{eqnarray}&&
p^0_l(\bm x', \bm x)=\frac{24 z(z-h)\rho_l}{h^3 \rho^2}=\frac{24z(z-h)}{ h^3}\nabla_l\ln \rho.
\end{eqnarray}
Here, $p^0$ is the leading order approximation for the pressure of the Stokeslet at large distances \cite{LironMochon}, with an exponentially small correction.
We can write ($l\neq z$ but $i$ or $k$ can be  $z$),
\begin{eqnarray}&&\!\!\!\!\!\!\!\!\!\!
T_{ilk}(\bm x', \bm x)=-p^0_l\delta_{ik}+\nabla'_k S^0_{il}+\nabla'_i S^0_{kl}\!+\!{\tilde T}_{ilk}(\bm x', \bm x),
\end{eqnarray}
where ${\tilde T}_{ilk}(\bm x', \bm x)$ decays exponentially in $\rho$ with exponent at least $\pi/h$, cf.\ Eq.~(\ref{st}). The stress tensor $T_{ilk}(\bm x', \bm x)$
is exponentially small when one of the indices is $z$.
We find using the expressions for $p_0$ and $S^0_{il}$,
\begin{eqnarray}&&
T_{ilk}(\bm x', \bm x)\approx \frac{24z(h-z)}{h^3}\nabla_l\left[ \left(\delta_{ik}+z'(h\!-\!z')\nabla_i \nabla_k \right)\ln \rho\right]\nonumber\\&&
+\frac{12 (h\!-\!2z')z(z\!-\!h)}{h^3}\left(\delta_{kz}\nabla_i+\delta_{iz}\nabla_k \right) \nabla_l\ln \rho,
\end{eqnarray}
where we neglected exponentially small correction. We find from Eq.~(\ref{inrep}) that,
\begin{eqnarray}&&\!\!\!\!\!\!\!\!\!\!
\bm u(\bm x)\!=\!\frac{z(z\!-\!h)\nabla p}{2\eta}+O\left(e^{-\pi\min[\rho_n]/h}\right)\!;\ \ \nabla^2 p\!=\!0,\label{fi}
\end{eqnarray}
where
$\bm u=(u_x, u_y)$ and $\min[\rho_n]$ is the distance from $\bm x$ to the closest boundary of a particle.
The pressure $p$ in this formula is independent of $z$ as predicted by the lubrication theory used in Eq.~(\ref{lubr}) with,
\begin{eqnarray}&&\!\!\!\!\!\!
p=p^0+\sum_n \delta p_n,\ \ \delta p_n=\frac{3\nabla_i}{\pi h^3}\int_{S_n}z'(z'-h) \ln \rho \sigma_{ik}'dS_k'\nonumber
\\&&\!\!\!\!\!\!
-\frac{6\eta}{\pi h^3} \int_{S_n} dS'_k  \left(\delta_{ik}+z'(h\!-\!z')\nabla_i \nabla_k \right) u_{i}(\bm x') \ln \rho\nonumber\\&&\!\!\!\!\!\!
+\frac{3\eta}{\pi h^3} \int_{S_n} dS'_k  (h\!-\!2z')\left(\delta_{kz}\nabla_i+\delta_{iz}\nabla_k \right) u_{i}(\bm x') \ln \rho,\label{fu}
\end{eqnarray}
where $\delta p_n$ is the pressure perturbation due to the $n-$th particle and the summation over repeated indices is from $1$ to $3$. Eqs.~(\ref{fi})-(\ref{fu}) are one of the main results of our work. These provide a refinement of the lubrication theory demonstrating that Eq.~(\ref{lubr}) holds under
the condition $\exp{(-\pi \min[\rho_n]/h)}\ll 1$, which is difficult to show
using the classic lubrication theory as it demands the strong inequality $\min[\rho_n]\gg h$. For instance, at $\min[\rho_n]=h$ the exponential factor is $\sim 0.04$. The result holds both for droplets and rigid particles where for rigid particles the last two lines of Eq.~(\ref{fu}) become zero and the equation reduces to
\begin{eqnarray}&&\!\!\!\!\!\!
p=p^0+\frac{3\nabla_i}{\pi h^3}\sum_n\int_{S_n}z'(z'-h) \ln \rho \sigma_{ik}'dS_k'.\label{rg}
\end{eqnarray}
We have good control of the correction terms to Eqs.~(\ref{fi})-(\ref{rg}) from the series representation of ${\tilde S}_{il}(\bm x', \bm x)$ provided in \cite{LironMochon}.

The pressure $p$ solves the two-dimensional Laplace equation in the domain between the particles since it is formed by integrals of the fundamental solution of the Laplace equation $\ln \rho$ over the particle boundaries.
The formula for $p$ matches the ideal flow that holds beyond the horizontal distance $h$ from the particles with the fully viscous flow near the particles. The viscous layer is the neighborhood
of the boundary of each particle where Eq.~(\ref{fu}) breaks down. Though the solution for $p$ is given in terms
of the unknown velocities and stress tensors on the surfaces of the particles, it seems that this is as much as can be done generally: the matching problem is not solvable for any general particle shape. It does simplify for disk-like particles as described previously.

{\it Hydrodynamic interactions of pancake-like droplets.}---Eqs.~(\ref{fi})-(\ref{fu}) provide support for the observation that  the width of the viscous layer around disk-like droplets is not larger than $h$.
The formulae tell that, unless the distance between the droplet surfaces is smaller than $h$, the (horizontal) flow outside the viscous layers near the particles is an
ideal potential flow. This flow can be determined using the boundary condition that the normal velocity at the outer boundary of the viscous layer coincides with the normal component of the translational velocity of the particle.
Since the layer width is of the order of $h$, and as long as the distance between the droplets is larger than $h$ (but possibly much smaller than $a$) we can impose the boundary condition on the particle surface, neglecting the finite width of this viscous layer as we did for the case of the single large disk, see Eq.~(\ref{dl}). Similarly, in the presence of many particles whose separation is larger than $h$, the flow outside the boundary layer is described by a pressure field $p$ that obeys \cite{is,tl2012},
\begin{eqnarray}&&
\nabla^2 p=0,\ \ \left(\bm v_n+\frac{h^2\nabla p}{12\eta}\right)\cdot{\hat n}_n=0, \label{press}
\end{eqnarray}
where $\bm v_n$ is the velocity of $n-$th particle, ${\hat n}_n$ is the unit vector normal to the surface of the $n-$th particle. The pressure gradient is taken at the outer boundary of the viscous layer of the $n-$th particle. However, since the latter is narrow, one can consider $\nabla p$ on the surface of the $n-$th particle without affecting significantly the solution for the pressure outside the viscous layers. To find the pressure inside the layers would require a separate study. For close droplets the pressure determined by Eq.~(\ref{press}) is different from the superposition of the dipole solutions given by Eqs.~(\ref{dlp0})-(\ref{dl}) due to the near-field interactions.

Finally, we demonstrate that the force exerted on the particles, determined by the viscous stress tensor at the particle surface, can be obtained from the ideal flow description.
The force $\bm F^n$ on particle $n$ is determined by the following integral over the particle surface,
\begin{eqnarray}&&
\bm F^n_i\!= \int_{S_n}\sigma_{ik}dS_k=\int_{outer}\sigma_{ik}dS_k,
\end{eqnarray}
where the last integral is over the outer boundary of the viscous layer of the $n-$th particle and we used $\nabla_k\sigma_{ik}=0$. We can neglect the viscous contribution to the stress tensor at the outer boundary and find
\begin{eqnarray}&&
\bm F^n\!\approx -\int_{outer}p d\bm S\approx -\int_{S_n}p d\bm S,\label{force}
\end{eqnarray}
where we must use the pressure $p$ determined from Eq.~(\ref{press}) in the last term and not the true pressure on the surface of the particle. Thus, the force
coincides with that in an ideal flow and, effectively, we can assume that the ideal flow holds everywhere disregarding the no-slip boundary condition. This provides a consistent basis for the study of hydrodynamic interactions between  large droplets at small distances as performed in \cite{is,tl2012}.


\section{Multipole expansion}

The flow at large distances from the particles can be  effectively studied using the multipole expansion. The distances must be larger than $h$ and much larger than the particle size.
We perform here this expansion in terms of $\delta p_n$ in Eq.~(\ref{fu}), solution of the two-dimensional
Laplace equation. We write $\delta p_n$ as,
\begin{eqnarray}&&\!\!\!\!\!\!\!\!\!
\delta p_n=\frac{3}{\pi h^3}\int_{S_n}dS_k' \left(
z'(z'-h) \sigma_{ik}'\nabla_i-\eta u_{i}(\bm x')\left(
2\delta_{ik}\nonumber\right.\right.\\&&\!\!\!\!\!\!\!\!\!\left.\left.+2z'(h\!-\!z')\nabla_i \nabla_k
+(h\!-\!2z')
\left(\delta_{kz}\nabla_i+\delta_{iz}\nabla_k \right)\right)\right)\ln \rho.\label{deltap}
\end{eqnarray}
We provide next the expansion
in Cartesian and polar coordinates
as in three-dimensional electrostatics \cite{cl}.

We set the origin of the coordinate system
inside the $n-$th particle. To determine the multipole expansion in Cartesian coordinates we consider the Taylor series (remind that $\rho=|\bm r-\bm r'|$),
\begin{eqnarray}&&\!\!\!\!\!\!
\ln \rho\!=\!\ln r\!-\!r'_l\nabla_l\ln r\!+\!\frac{r'_lr'_p}{2}\nabla_l\nabla_p\ln r\!+\!\ldots,
\end{eqnarray}
where dots stand for higher-order terms. Substituting into Eq.~(\ref{deltap}) one obtains the Cartesian form of the multipole expansion. The  leading-order $\ln r$ term in the series,
\begin{eqnarray}&&\!\!\!\!\!\!
\delta p_n=\frac{3}{\pi h^3}\int_{S_n}dS_k' \left(
z'(z'-h) \sigma_{ik}'\nabla_i-\eta u_{i}(\bm x')\left(
2\delta_{ik}\nonumber\right.\right.\\&&\!\!\!\!\!\!\left.\left.+2z'(h\!-\!z')\nabla_i \nabla_k
+(h\!-\!2z')
\left(\delta_{kz}\nabla_i+\delta_{iz}\nabla_k \right)\right)\right)\ln r,
\end{eqnarray}
has a contribution proportional to $\int \bm u\cdot{\bm dS}$, i.e.\  proportional to $\ln r$. Further assuming the droplet is incompressible $\int \bm u\cdot{\bm dS}=0$.
In this case,  the leading order term at larger distances is given by the dipole term,
\begin{eqnarray}&&\!\!\!\!\!\!
\delta p_n=\frac{3}{\pi h^3}\int_{S_n}dS_k' \left(
z'(z'-h) \sigma_{ik}'\nabla_i-\eta u_{i}(\bm x')(h\!-\!2z')\nonumber\right.\\&&\!\!\!\!\!\!\left.
\left(\delta_{kz}\nabla_i+\delta_{iz}\nabla_k \right)\right)\ln r
+\frac{6\eta\nabla_l\ln r}{\pi h^3}\int_{S_n}r'_l \bm u\cdot{\bm dS}, \label{dpl}
\end{eqnarray}
where the last term comes from the next-order term in the expansion of the logarithm. This term can be simplified for droplets that do not change their shape, such as the pancake-like droplets considered previously, since
the slip and flow on the surface are irrelevant. For instance, for a spherical droplet whose center moves with velocity $\bm v$,  one obtains $\int_{S_n}r'_l \bm u\cdot{\bm dS}=v_k\int_{S_n}r'_l dS_k=4\pi a^3 v_l/3$.
The complete expansion becomes,
\begin{eqnarray}&&\!\!\!\!\!\!\!\!\!
\delta p_n=\frac{3}{\pi h^3}\int_{S_n}dS_k' \left(
z'(z'-h) \sigma_{ik}'\nabla_i-\eta u_{i}(\bm x')\left(
2\delta_{ik}\nonumber\right.\right.\\&&\!\!\!\!\!\!\!\!\!\left.\left.+2z'(h\!-\!z')\nabla_i \nabla_k
+(h\!-\!2z')
\left(\delta_{kz}\nabla_i+\delta_{iz}\nabla_k \right)\right)\right)\left(\ln r\nonumber\right.\\&&\!\!\!\!\!\!\left.
-r'_{l}\nabla_{l}\ln r+\frac{r'_{l}r'_{p}}{2}\nabla_{l}\nabla_{p}\ln r+\ldots
\right).
\end{eqnarray}
The expansion in polar coordinates is found observing that for $r'<r$,
\begin{eqnarray}&&
\ln |\bm r-\bm r'|=\ln r\nonumber\\&&
-\sum_{n=1}^{\infty}\left(\frac{r'}{r}\right)^n\frac{\cos(n\theta)\cos(n\theta')+\sin(n\theta)\sin(n\theta')}{n}. \label{logexp}
\end{eqnarray}
This formula represents the fundamental solution $\ln |\bm r-\bm r'|$ in terms of the elementary solutions of Laplace equation, $r^{-k}\exp(ik\theta)$ and $r'^p\exp(ip\theta')$ with  $k$ and $p$ positive integers.
This is the counterpart of the expansion of $|\bm r-\bm r'|^{-1}$ in spherical harmonics adopted in three-dimensional multipole expansion in electrostatics \cite{cl} and, in fact, it can be derived from
that expansion by confining $\bm r$, $\bm r'$ in a plane. We provide here a simpler derivation. We consider,
\begin{eqnarray}&&\!\!\!\!\!\!
\ln |\bm r-\bm r'|=\ln r+\frac{\ln\left(1-2\epsilon\cos\gamma+\epsilon^2\right)}{2},\ \ \epsilon=\frac{r'}{r},
\end{eqnarray}
where $\gamma$ is the angle between $\bm r$ and $\bm r'$ and $\epsilon<1$. We recall the Fourier series,
\begin{eqnarray}&&
\ln\left(1\!-\!2\epsilon\cos\gamma\!+\!\epsilon^2\right)\!=\! -\sum_{n=1}^{\infty}\frac{2\epsilon^n \cos(n\gamma)}{n}\,
\end{eqnarray}
where the integrals for the Fourier coefficients can be obtained using the residue theorem \cite{grad}.
Finally, introducing the polar angles $\theta$ and $\theta'$ for $\bm r$ and $\bm r'$, respectively, and using $\gamma=\theta'-\theta$ we obtain Eq.~(\ref{logexp}).
The multipolar expansion in polar coordinates is finally
\begin{eqnarray}&&\!\!\!\!\!\!\!\!\!
\delta p_n=\frac{3}{\pi h^3}\int_{S_n}dS_k' \left(
z'(z'-h) \sigma_{ik}'\nabla_i-\eta u_{i}(\bm x')\left(
2\delta_{ik}\nonumber\right.\right.\\&&\!\!\!\!\!\!\!\!\!\left.\left.+2z'(h\!-\!z')\nabla_i \nabla_k
+(h\!-\!2z')
\left(\delta_{kz}\nabla_i+\delta_{iz}\nabla_k \right)\right)\right)\left(\ln r\nonumber\right.\\&&\!\!\!\!\!\!\left.
-\sum_{n=1}^{\infty}\left(\frac{r'}{r}\right)^n\frac{\cos(n\theta)\cos(n\theta')+\sin(n\theta)\sin(n\theta')}{n}
\right),
\end{eqnarray}
which gives the pressure as a superposition of elementary solutions $r^{-k}\exp(ik\theta)$.
The formulae simplify for rigid particles to
\begin{eqnarray}&&\!\!\!\!\!\!\!\!\!
\delta p_n=\frac{3}{\pi h^3}\int_{S_n}dS_k'
z'(z'-h) \sigma_{ik}'\nabla_i\left(\ln r\nonumber\right.\\&&\!\!\!\!\!\!\left.
-r'_{l}\nabla_{l}\ln r+\frac{r'_{l}r'_{p}}{2}\nabla_{l}\nabla_{p}\ln r+\ldots
\right)
\end{eqnarray}
and
\begin{eqnarray}&&\!\!\!\!\!\!\!\!\!
\delta p_n=\frac{3}{\pi h^3}\int_{S_n}dS_k'
z'(z'-h) \sigma_{ik}'\nabla_i\left(\ln r\nonumber\right.\\&&\!\!\!\!\!\!\left.
-\sum_{n=1}^{\infty}\left(\frac{r'}{r}\right)^n\frac{\cos(n\theta)\cos(n\theta')+\sin(n\theta)\sin(n\theta')}{n}
\right).
\end{eqnarray}
The multipole expansion of the flow is derived by taking the gradient of the pressure using Eq.~(\ref{fi}). As an example, the perturbation  of the Poiseuille flow due to a rigid particle, $\delta u_k$, is up to a cubically decaying term,
\begin{eqnarray}&&
\delta u_k\!
=\!-\frac{3z(z\!-\!h)s_i}{2\pi h^3\eta}\nabla_k \nabla_i \ln r -\frac{3z(z\!-\!h)}{2\pi h^3\eta}\nonumber \\&&\times\nabla_k \nabla_i\frac{1}{r}\int_{S}\!\!\!dS_k'z'(z'\!-\!h) \sigma_{ik}' r' \cos(\theta-\theta'),\label{mo}
\end{eqnarray}
where we introduced,
\begin{eqnarray}&&\!\!\!\!\!\!\!\!\!\!\!\!\!\!\!\!
s_i=\int_{S}\!\! z'(h\!-\!z') \sigma_{ik}'dS_k'. \label{dlpa}
\end{eqnarray}
The first term in Eq.~(\ref{mo}) is a dipole and the second term a quadrupole. Similarly, we can write the corresponding and higher-order terms for droplets.

\section{Leading-order behavior at large distances}

In this Section we consider the leading order behavior of the flow at large horizontal distances from the particle(s). The distances must be larger than $h$ (but not much larger) and much larger than the size of the particles.
The far-field flow perturbation $\delta \bm u_n$ due to the $n-$th particle is given by the dipole flow,
\begin{eqnarray}&&\!\!\!\!\!\!
\delta \bm u_n=\frac{z(z\!-\!h)}{2\eta}\nabla \delta p_n,\ \ \delta p_n=-\frac{3}{\pi h^3}\!(\bm s_n\cdot\nabla) \ln r,\label{flowdp}
\end{eqnarray}
where
\begin{eqnarray}&&\!\!\!\!\!\!
(s_n)_i=\int_{S_n}dS_k'
z'(h-z') \sigma_{ik}'+\eta\int_{S_n}dS_z' (h\!-\!2z')u_{i}(\bm x')\nonumber\\&&\!\!\!\!\!\!+\eta\int_{S_n}dS_i' (h\!-\!2z')u_{z}(\bm x')
-2\eta\int_{S_n}r'_i \bm u\cdot{\bm dS},
\end{eqnarray}
see Eqs.~(\ref{fi}) and (\ref{dpl}). For rigid particles this reduces to Eq.~(\ref{dlpa}) which is why we use the same letter for the coefficient $s_i$. We find that the perturbation of the potential of the
depth-averaged flow $\delta \phi_n=-h^2 \delta p_n/(12\eta)$ is,
\begin{eqnarray}&&\!\!\!\!\!\!\!\!\!\!\!\!\!
\delta \phi_n=-\frac{\tilde {\bm d}_n\cdot{\hat r}}{r},\ \ \tilde {\bm d}_n=-\frac{\bm s_n }{4\pi\eta h}, \label{tld}
\end{eqnarray}
see Eq.~(\ref{dlp0}). Thus, the flow perturbation at large distances is the dipolar flow with effective dipole moment $\tilde {\bm d}$.
We can describe the far-field impact of a particle of arbitrary shape on the flow introducing
the source in the potential equation, $\nabla^2\delta\phi_n=-2\pi(\tilde {\bm d}_n\cdot\nabla)\delta(x)\delta(y)$ so that the full potential $\phi$ obeys,
\begin{eqnarray}&&\!\!\!\!\!\!\!\!\!\!\!\!\!
\nabla^2\phi=-2\pi\sum_n(\tilde {\bm d}_n\cdot\nabla)\delta(x-x_n)\delta(y-y_n)
\end{eqnarray}
where $(x_n, y_n)$ are the horizontal coordinates of some point inside the $n-$th particle (observe that $p^0$ is a linear function and has zero laplacian).

The resulting correction to the Poiseuille flow is that of
a particle that moves in direction of $\bm s$, see Eq.~(\ref{dl}),
\begin{eqnarray}&&\!\!\!\!\!\!\!\!\!\!\!\!\!\!\!\!
\delta u_k(\bm x)\!=\!\frac{3s_iz(h\!-\!z)}{\pi \eta h^3(x^2+y^2)}\left[\frac{\delta_{ik}}{2}\!-\!\frac{x_ix_k}{x^2+y^2}\right],\ \ \rho\gg h. \label{stks}
\end{eqnarray}
The lateral, $y$ and $z$, components of $\bm s$ vanish for particles that have fore-and-aft symmetry. This can be shown in the same way as for the absence of lateral migration of spheres in a channel \cite{leal,1rd2}.
The reversal of the sign of $\bm u$ and $p$ produces another solution of the system of Eqs.~(\ref{ds}). This solution has opposite sign of the stress tensor and velocity and thus of $\bm s$. However, it describes
the same physical situation and thus must have the same lateral components of $\bm s$, hence these components must vanish. Thus, for spheres or ellipsoids $\bm s=s\bm {\hat x}$. In contrast, for particles whose shape
is an arc or similar one can have a non-zero $s_y$ and $s_z$.

We consider a spherical particle as an example of a particle with fore-and-aft symmetry. We can introduce $s_i=s(z_p)\delta_{ix}$ where,
\begin{eqnarray}&&\!\!\!\!\!\!
s(z_p)=\int_{S}\!\! z'(h\!-\!z') \sigma_{xk}'dS_k'+\eta\int_{S_n}dS_z' (h\!-\!2z')u_x(\bm x')\nonumber\\&&\!\!\!\!\!\!+\eta\int_{S_n}dS_x' (h\!-\!2z')u_{z}(\bm x')
-2\eta\int_{S_n}r'_x \bm u\cdot{\bm dS}, \ \
\end{eqnarray}
with $z_p$ the vertical position of the particle center. In this case, the flow is
\begin{eqnarray}&&\!\!\!\!\!\!\!\!\!\!\!\!\!\!\!\!
\delta u_x(\bm x)\!=\!\frac{3s(z_p)z(h\!-\!z)}{2\pi \eta h^3(x^2+y^2)}\frac{y^2- x^2}{x^2+y^2},\nonumber\\&&\!\!\!\!\!\!\!\!\!\!\!\!\!\!\!\!\!
\delta u_y(\bm x)\!=\!-\frac{3s(z_p)z(h\!-\!z)}{2\pi \eta h^3(x^2+y^2)}\frac{2xy}{x^2+y^2}.
\label{dist_flow}
\end{eqnarray}
It is plausible that $s(z_p)>0$ because the particle is always lagging behind the local flow. This is confirmed by the direct numerical simulations reported below.

The formulas provided here give the possibility of tabulating the particle dipole moments from numerical simulations for future use.
For a spherical particle of fixed radius the dipole moment depends on the vertical coordinate $z_p$.
The solution of the flow equations in the presence of an isolated sphere would give the stress tensor and the surface velocity
with which we can find $s(z_p)$. We illustrate this procedure for the case where the particle is a rigid sphere with same density as that of the fluid.
The equation of motion is,
\begin{equation}
    m\frac{dv_i}{dt} =
    \int_{S} \sigma_{ik} dS_k,
\label{eqmp}
\end{equation}
where $m$ is the mass of the particle and gravity does not influence the motion since the particle is assumed to be neutrally buoyant. This equation is coupled to the time-
dependent Navier-Stokes equations where the unsteady
term is not negligible during the transients.
The particle eventually reaches a constant velocity and the fluid flow is governed by the steady Stokes equations due to the small
Reynolds number.
Thus, our derivation of the far flow holds with the dipole coefficient for rigid particles,
\begin{eqnarray}&&\!\!\!\!\!\!\!\!\!\!\!\!\!\!\!\!
s(z_p)=\int_{S}\!\! z'(h\!-\!z') \sigma_{xk}'dS_k'.
\end{eqnarray}
We computed here $s(z_p)$ using the numerically determined $\sigma_{xk}$ for different $z_p$ and different radii of the sphere. The results are summarized in Tables \ref{tab: h=3} - \ref{tab: h=1.125} (see Appendix \ref{numm} for details). Here, the particle relative velocity, defined as $\delta u_p=u_p-u_x^0$ where $u_p$ is the particle center velocity, is non-dimensionalized by the bulk velocity of the undisturbed channel flow $U_b=-h^2\nabla_x p^0/(12\eta)$.
The reduction of the translational velocity as the particle is placed closer to one wall or as the particle size increases is consistent with previous computations using the
boundary integral method \cite{bim_1p}.

\begin{table}[t]
 \centering
   \caption{Relative particle velocity $\delta u_p/U_b$ and magnitude of $\hat{s}$ as function of the particle centre position $z_p/h$ for spherical particles of radius $a=h/6$ obtained from the numerical simulations. }
   \tabulinesep=1.2mm
   \begin{tabular}{l r r r r r r r}
       \hline
       $z_p/h$              &$0.50$   &$0.55$   &$0.60$   &$0.65$   &$0.70$  &$0.75$   &$0.80$\\
       \hline
       $\delta u_p/U_b$     &$-0.06$  &$-0.06$  &$-0.06$  &$-0.07$  &$-0.07$ &$-0.09$  &$-0.17$\\
       $\hat{s} \e{-3}$     &$1.8$    &$2.2$    &$3.7$    &$6.3$    &$10.6$   &$17.8$    &$30.0$\\
       \hline
   \end{tabular}
   \label{tab: h=3}
\end{table}

\begin{table}[t]
 \centering
   \caption{Relative particle velocity $\delta u_p/U_b$ and magnitude of $\hat{s}$ as function of the particle centre position $z_p/h$ for spherical particles of radius $a=h/3$ obtained from the numerical simulations. }
   \tabulinesep=1.2mm
   \begin{tabular}{l r r r r r}
       \hline
       $z_p/h$              &$0.50$   &$0.55$   &$0.60$    &$0.65$\\
       \hline
       $\delta u_p/U_b$     &$-0.24$  &$-0.25$  &$-0.30$   &$-0.45$\\
       $\hat{s} \e{-2}$     &$4.1$    &$4.9$    &$7.6$     &$15.8$\\
       \hline
   \end{tabular}
   \label{tab: h=1.5}
\end{table}

\begin{table}[t!]
 \centering
   \caption{Relative particle velocity $\delta u_p/U_b$ and magnitude of $\hat{s}$ as function of the particle centre position $z_p/h$ for spherical particles of radius $a=h/2.25$ obtained from the numerical simulations. }
   \tabulinesep=1.2mm
   \begin{tabular}{l r}
       \hline
       $z_p/h$              &$0.50$\\
       \hline
       $\delta u_p/U_b$     &$-0.52$\\
       $\hat{s} \e{-1}$     &$2.1$\\
       \hline
   \end{tabular}
   \label{tab: h=1.125}
\end{table}

\begin{figure}[t!]
 \begin{center}
\includegraphics[width=.9\columnwidth]{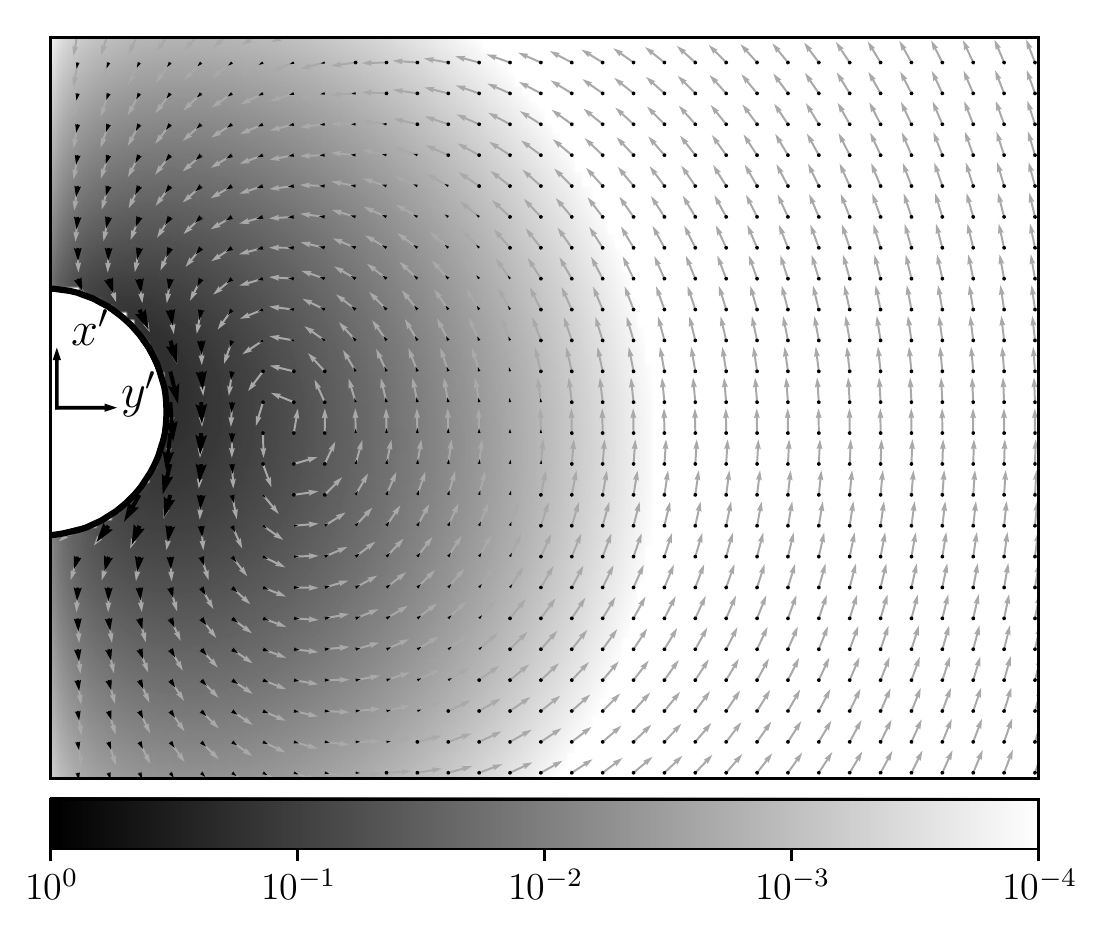}
\end{center}
  \caption{Depth-averaged disturbance flow around a sphere from the numerical simulations. The heavy arrows indicate the magnitude and light arrows the direction. The color in the background depicts the decay of the vorticity outside the sphere. The particle travels in the $x'$ direction and is located at the mid-channel ($z_p/h=0.5$), with $h/(2a)=1.5$. Only half of the plane is shown due to symmetry.}
 \label{fig: flow}
\end{figure}

We also report the values of
$\hat{s}(z_p)=3s(z_p)/(2\pi\eta h^3)$, 
the common pre-factor in Eq.~(\ref{dist_flow}). We note that, though overall small, $\hat{s}$ increases as the particle approaches one wall or as the confinement increases, as the relative velocity $\delta u_p$. The resulting $\hat{s}$, quantifying the local velocity disturbance generated by one particle, along with the spatial dependence in the horizontal plane, allows for predictions of the far-field interactions of spheres. These will be examined  in Section \ref{interactions}.

\begin{figure}[t!]
 \begin{center}
 \includegraphics[width=.9\columnwidth]{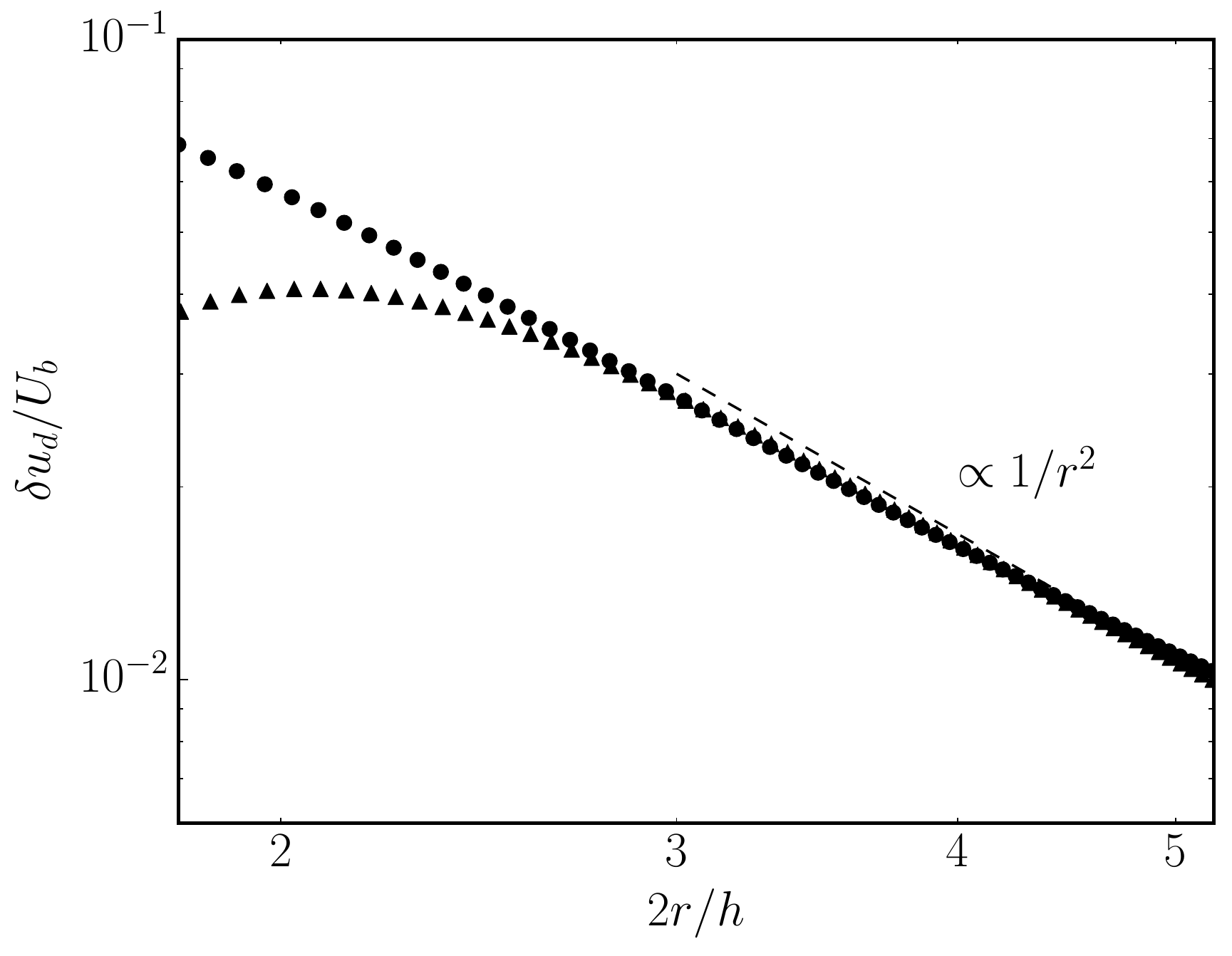}
 \end{center}
 \caption{Spatial variation of the normalized streamwise depth-averaged disturbance-velocity, $\delta u_d/U_b$, along the streamwise ($y=0$, circle) and spanwise ($x=0$, triangle) directions away from the particle center. The particle is located at the mid-channel ($z_p/h=0.5$), with $h/(2a)=1.125$.
 The collapse of the disturbance-velocity away from the particle confirms the leading-order dipolar decay (dashed line).}
 \label{fig: decay}
\end{figure}

Finally, we return to the lubrication theory by showing some typical depth-average velocity field in Fig.\ \ref{fig: flow} and the velocity decay in Fig.\ \ref{fig: decay}. As mentioned earlier,
the lubrication theory is valid at horizontal distances larger than the height of the channel.
Fig.\ \ref{fig: flow} depicts the flow field due to a sphere of diameter equal to 2/3 of the channel height. The non-zero vertical vorticity outside the particle indicates the non-ideal structure of the depth-average flow, in contrast to the simple mass dipole of a disk (see Eq.~\ref{dl}). However, as the confinement increases, the disturbance velocity asymptotes the leading-order quadratic decay, as shown in Fig.\ \ref{fig: decay}.

\section{Interactions}\label{interactions}

In this Section, we introduce equations that describe interactions of well-separated particles and solve them in some specific cases. We start by observing that because of the
linearity of the problem, the steady state (horizontal) velocity $\bm v_0$ of an isolated particle driven by the Poiseuille flow according to Eq.~(\ref{ds}) is given by
\begin{eqnarray}&&\!\!\!\!\!\!\!\!\!\!
\bm v_0=-{\hat M}\nabla p^0, \label{steady}
\end{eqnarray}
where we assume $\bm v_0$ is a function of $\nabla p^0$ that can be any constant vector in the plane. Indeed, $\bm v_0$ is a linear function of $\nabla p^0$ that is zero when there is no driving flow. Since ${\hat M}$  connects the velocity $\bm v_0$ with the force per unit volume of the fluid,  we call ${\hat M}$ the mobility matrix though it differs
from the more commonly used coefficient between the velocity and the force on the particle \cite{hb}.
The two-by-two mobility matrix ${\hat M}$ depends on the shape of the particle, whether the particle is rigid or droplet, and the particle position in the channel. The equation neglects gravitational settling, absent for neutrally buoyant particles or particles whose sedimentation is stopped by interactions with the walls,
(as the pancake-like droplets) or because settling is negligible at relevant time scales. In cases with sedimentation velocity $\bm v_s$ so low that the particle stays in quasi-steady state we have,
\begin{eqnarray}&&\!\!\!\!\!\!\!\!\!\!
\bm v_0=-{\hat M}(t)\nabla p^0+\bm v_s,
\end{eqnarray}
where the matrix ${\hat M}(t)$ is determined by the instantaneous configuration in the channel, which may depend on time due to sedimentation.

We next consider interactions of many well-separated particles. The flow induced at the position of the $i-$th particle by the other particles is a quasi-Poiseuille flow,
\begin{eqnarray}&&\!\!\!\!\!\!\!\!\!\!
\bm u\!=\!\frac{z(z\!-\!h)\nabla p}{2\eta},\ \ p\!=\!p^0\!-\!\sum_{k\neq i} \frac{3}{\pi h^3}\!(\bm s_k\cdot\nabla) \ln |\bm r\!-\!\bm r_k|,
\end{eqnarray}
where $\bm r_k$ is the horizontal position of the $k-$th particle, see Eqs.~(\ref{fi}), (\ref{flowdp}). We observe that we can neglect variations of $\nabla p$ over the particle since  the rest of the particles are well-separated. Thus, at the leading order in large distances between the particles the $i-$th particle assumes the horizontal velocity,
\begin{eqnarray}&&\!\!\!\!\!\!\!\!\!\!
\frac{d\bm r_i}{dt}
\!=\!-{\hat M}_i\nabla\left( p^0\!-\!\sum_{k\neq i} \frac{3}{\pi h^3}\!(\bm s_k\cdot\nabla) \ln |\bm r\!-\!\bm r_k|\right)_{\bm r=\bm r_i}, \label{inhetractio}
\end{eqnarray}
where ${\hat M}_i$ describes the geometry of the $i-$th particle. The sedimentation velocity can be included in a straightforward way. This is the equation that describes the long-range interactions of the particles. The presented derivation avoids the problem with boundary conditions encountered in the derivation of \cite{tl2014} for the case of droplets. In that case, the derivation started with the flow induced by other particles at the position of the $i-$th particle and not the pressure. Since for particles of finite extent it becomes non-obvious where the three-dimensional flow must be considered, our derivation seems to be useful for a proper consideration of particles whose vertical size is smaller than $h$.

We consider the case of spherical particles or droplets of radius $a$ smaller than $h/2$. In this case ${\hat M}_i$ is $M(z_i)$ times the unit matrix where the scalar coefficient $M$ depends on the vertical coordinate $z_i$ of the $i-$th particle. Similarly $\bm s_k=s(z_k){\hat x}$ where $s(z)$ was introduced previously. We find,
\begin{eqnarray}&&
\dot{\bm r_i}
\!=\!-\! M(z_i)\nabla p^0\! +\!\sum_{k\neq i}\! \frac{3M(z_i) s(z_k)}{\pi h^3r_{ik}^2}\!\left[\bm {\hat x}\!-\!\frac{2 \left(\bm r_{ik}\cdot\bm {\hat x}\right) \bm r_{ik}}{r_{ik}^2}\right], \nonumber
\end{eqnarray}
where $\bm r_{ik}=\bm r_i-\bm r_k$. Thus for pair of particles,
\begin{eqnarray}&&
\dot{\bm r}
\!=
\left(M(z_1)\!-\! M(z_2)\right)
\nabla p^0\! +\!\frac{3\delta_{12}}{\pi h^3r^2}\!\left[\bm {\hat x}\!-\!\frac{2 x \bm r}{r^2}\right], 
\label{pair}
\end{eqnarray}
where $\bm r=\bm r_2-\bm r_1=(x, y, z)$ and we introduced,
\begin{eqnarray}&&
\delta_{12}=M(z_2) s(z_1)-M(z_1) s(z_2).
\end{eqnarray}
Another case when Eqs.~(\ref{inhetractio}) simplify significantly is for pancake-like droplets that almost completely fill the channel in the vertical direction. In this case $M$ and $s$ are constant since no variation of the vertical position of the particles is possible. We see immediately that the configuration of two droplets is stable in the dipole approximation where $\dot{\bm r}$ in Eq.~(\ref{pair}) is zero (in a higher-order quadrupole approximation proportional to $r^{-3}$ the pair would not be stable). For many particles, the equations of motion in the frame that moves with the velocity of the isolated droplet $-M\nabla p^0$ become,
\begin{eqnarray}&&\!\!\!\!\!\!\!\!\!\!\!\!\!\!\!\!
\frac{d\bm r_i}{dt}=\sum_{k\neq i}\frac{q}{r_{ik}^2}\left[\bm {\hat x}\!-\!\frac{2 \left(\bm r_{ik}\cdot\bm {\hat x}\right) \bm r_{ik}}{r_{ik}^2}\right],\ \ q= \frac{3M s}{\pi h^3}. \label{ints}
\end{eqnarray}
These equations hold also for spherical particles located at the same distance from the mid-plane where we must use for $M$ and $s$ the values at the corresponding $z$. This is the case where the particles have identical vertical coordinate or their coordinates can be obtained by reflection with respect to the mid plane. Other cases of symmetric particles where Eq.~(\ref{ints}) hold can be considered. If gravitational settling is relevant,  $\bm r_{ik}$ will change via time-dependent $s=s(z(t))$. It is assumed below that the change of $s=s(z(t))$ can be neglected over the time scales of interest.

%


It is often the case that we have two spherical particles at the same vertical distance from the walls. This can be the case of spherical droplets created at some fixed place in the channel and then transported down the flow \cite{tab0}. In this case, an isolated pair is stable in the dipole approximation: we have $\dot{\bm r}=0$ in Eq.~(\ref{pair}) for $z_1=z_2$.
This characterizes the basic property of the interaction given by Eq.~(\ref{ints}), that the velocity induced by particle $i$ at the position of the $k-$th particle is equal to the velocity induced by particle $k$ at the position of the $i-$th particle. Thus the interparticle distances can change only if there are three or more particles. We can re-write the equation of motion as
\begin{eqnarray}&&\!\!\!\!\!\!\!\!\!\!\!\!\!\!\!\!
\frac{d\bm r_{ik}}{dt}=\sum_{l\neq i, l\neq k}\left(\frac{q}{r_{il}^2}\left[\bm {\hat x}\!-\!\frac{2 \left(\bm r_{il}\cdot\bm {\hat x}\right) \bm r_{il}}{r_{il}^2}\right]\right.\nonumber\\&&\!\!\!\!\!\!\!\!\!\!\!\!\!\!\!\!\left.-\frac{q}{r_{kl}^2}\left[\bm {\hat x}\!-\!\frac{2 \left(\bm r_{kl}\cdot\bm {\hat x}\right) \bm r_{kl}}{r_{kl}^2}\right]\right).
\end{eqnarray}
We start by considering in more detail the simplest case of two particles whose distance is constant in time. If the particles have the same $y-$coordinate then the $x-$coordinates obey,
\begin{eqnarray}&&\!\!\!\!\!\!\!\!\!\!\!\!\!\!\!\!
\frac{dx_1}{dt}=\frac{dx_2}{dt}=-\frac{q}{(x_1-x_2)^2}.
\end{eqnarray}
In this case the particles form a simple cluster with fixed distance that  moves as a whole slower than the particles separately. We consider now two particles at different spanwise locations, $y_1=y$, $y_2=0$,
\begin{eqnarray}&&\!\!\!\!\!\!\!\!\!\!\!\!\!\!\!\!
\frac{dx_1}{dt}=\frac{dx_2}{dt}=\!q\frac{(y_1-y_2)^2-(x_1-x_2)^2}{\left[(x_1-x_2)^2+(y_1-y_2)^2\right]^2},\nonumber\\&&\!\!\!\!\!\!\!\!\!\!\!\!\!\!\!\!\!
\frac{dy_1}{dt}=\frac{dy_2}{dt}\!=\!-q\frac{2(x_1-x_2)(y_1-y_2)}{\left[(x_1-x_2)^2+(y_1-y_2)^2\right]^2}.
\end{eqnarray}
The RHSs are constant because inter-particle distances are but the velocity of the cluster of the two particles can change sign unlike the previous case. For two particles with the same $x$ coordinate, the $x-$ component of their velocity increases while the $y-$component is zero, see \cite{tab0} for experimental observations.

Next, we consider the simplest case with changing inter-particle distances: three particles at the same height. From the analysis of the two-particle dynamics, a possible solution is that particles form a cluster of two particles with the third farther away.
%
The interactions of the single distant particle with the clustered particles decay quadratically with the distance and can be assumed negligible. Thus, the isolated particle moves with the velocity of one single sphere. The cluster keeps its configuration and moves at a constant velocity $(q/r^2)[\bm {\hat x}\!-\!2 \left(\bm r\cdot\bm {\hat x}\right) \bm r/r^2]$ with respect to the third particle. If this velocity is such to increase the separation between the cluster and the third particle, this solution will continue ad infinitum. It is thus plausible to assume that any arbitrary initial configuration of three particles will separate asymptotically in one cluster and one particle. We will prove this below for the practically important case of three particles aligned in the streamwise, $x-$, direction. This case can be observed when the particles are injected in the flow at the same location.

The distances between three particles are determined by two vectors $\bm r_{12}$ and $\bm r_{13}$ that obey,
\begin{eqnarray}&&\!\!\!\!\!\!\!\!
\dot{\bm r}_{12}\!=\!\frac{q}{r_{13}^2}\left[\bm {\hat x}\!-\!\frac{2 \left(\bm r_{13}\!\cdot\!\bm {\hat x}\right) \bm r_{13}}{r_{13}^2}\right]\!-\!\frac{q}{r_{23}^2}\left[\bm {\hat x}\!-\!\frac{2 \left(\bm r_{23}\!\cdot\!\bm {\hat x}\right) \bm r_{23}}{r_{23}^2}\right],\nonumber\\&&\!\!\!\!\!\!\!\!
\dot{\bm r}_{13}\! =\!\frac{q}{r_{12}^2}\left[\bm {\hat x}\!-\!\frac{2 \left(\bm r_{12}\!\cdot\!\bm {\hat x}\right) \bm r_{12}}{r_{12}^2}\right]\!-\!\frac{q}{r_{23}^2}\left[\bm {\hat x}\!-\!\frac{2 \left(\bm r_{23}\!\cdot\!\bm {\hat x}\right) \bm r_{23}}{r_{23}^2}\right]\!.\label{grv}
\end{eqnarray}
The  solution described above pertaining the cluster of two particles (named here $2$ and $3$) and the faraway particle $1$ corresponds to neglecting the first terms in the RHSs,
\begin{eqnarray}&&\!\!\!\!\!\!\!\!\!\!\!\!\!\!\!\!
\dot{\bm r}_{12}\approx \dot{\bm r}_{13}\approx -\frac{q}{r_{23}^2}\left[\bm {\hat x}\!-\!\frac{2 \left(\bm r_{23}\!\cdot\!\bm {\hat x}\right) \bm r_{23}}{r_{23}^2}\right]\approx const,\label{clust}
\end{eqnarray}
where $\bm r_{23}$ is approximately constant. At large times the constant vector $\bm r_{23}$ has become much smaller than the linearly growing ${\bm r}_{12}$ and ${\bm r}_{13}$. We have that $\bm r_{23}={\bm r}_{13}-{\bm r}_{12}$ obeys the equation,
\begin{eqnarray}&&\!\!\!\!\!\!\!\!
\dot{\bm r}_{23}\!=\!\frac{q}{r_{12}^2}\left[\bm {\hat x}\!-\!\frac{2 \left(\bm r_{12}\!\cdot\!\bm {\hat x}\right) \bm r_{12}}{r_{12}^2}\right]\!-\!\frac{q}{r_{13}^2}\left[\bm {\hat x}\!-\!\frac{2 \left(\bm r_{13}\!\cdot\!\bm {\hat x}\right) \bm r_{13}}{r_{13}^2}\right],\nonumber
\end{eqnarray}
where the RHS decays quadratically with time, in agreement with the assumption of constant  $\bm r_{23}$.

We prove that the separation of 3 particles into one binary cluster and one isolated particle holds for arbitrary initial conditions when all three particles lie on the same line in the $x-$direction. It is clear that the separation can occur in two ways in this case: either particles $1$ and $2$ form a cluster or $2$ and $3$. Here we assume the ordering $x_1>x_2>x_3$. As the cluster moves slower than the isolated particle, the third particle would catch up with the binary cluster, made of $1$ and $2$. Hence the only stable configuration is a cluster of particles $2$ and $3$ whose distance from particle $1$ increases linearly with time due to the cluster deceleration. Formally,
\begin{eqnarray}&&\!\!\!\!\!\!\!\!\!\!\!\!\!\!\!\!
\dot{x}_{12}\!=\!-q\left(\frac{1}{x_{13}^2}-\frac{1}{(x_{13}-x_{12})^2}\right),\nonumber\\&&\!\!\!\!\!\!\!\!\!\!\!\!\!\!\!\!\!
\dot{x}_{13}\!=\!-q\left(\frac{1}{x_{12}^2}-\frac{1}{(x_{13}-x_{12})^2}\right).\label{s}
\nonumber
\end{eqnarray}
Introducing $x=x_{12}$ and $r=x_{13}/x_{12}$ where $r>1$, we can write
\begin{eqnarray}&&\!\!\!\!\!\!\!\!\!\!\!\!\!\!\!\!
\dot{x}\!=\!-\frac{q}{x^2}\left(\frac{1}{r^2}-\frac{1}{(r-1)^2}\right)\!=\!-\frac{q\left(1-2r\right)}{x^2 r^2(r-1)^2},\nonumber\\&&\!\!\!\!\!\!\!\!\!\!\!\!\!\!\!\!\!
\dot{x}+\frac{x \dot{r}}{r}\!=\!-\frac{q}{x^2}\left(\frac{1}{r}-\frac{1}{r(r-1)^2}\right).
\end{eqnarray}
The distance $r$ obeys,
\begin{eqnarray}&&\!\!\!\!\!\!\!\!\!\!\!\!\!\!\!\!
x^3\dot{r}\!=\!-q\left(1-\frac{1}{(r-1)^2}\right)+q\left(\frac{1}{r}-\frac{r}{(r-1)^2}\right)\nonumber\\&&\!\!\!\!\!\!\!\!\!\!\!\!\!\!\!\!\!
=-\frac{q(r-1)}{r}-\frac{q}{r-1}=-\frac{q(r^2\!-\!r\!+\!1)}{r(r-1)}.
\end{eqnarray}
Hence $r(t)$ decreases in time monotonously and we can write
\begin{eqnarray}&&\!\!\!\!\!\!\!\!\!\!\!\!\!\!\!\!
\frac{d\ln x}{ds}=q\left(\frac{1}{(r-1)^2}-\frac{1}{r^2}\right),\nonumber\\&&\!\!\!\!\!\!\!\!\!\!\!\!\!\!\!\!\!
\frac{dr}{ds}=-\frac{q(r^2\!-\!r\!+\!1)}{r(r-1)},\ \ \frac{ds}{dt}=\frac{1}{x^3(t)}.
\end{eqnarray}
We can solve for $s(r)$,

\begin{eqnarray}&&\!\!\!\!\!\!\!\!\!\!\!\!\!\!\!\!
\frac{ds}{dr}=-\frac{1}{q}+\frac{1}{q(r^2\!-\!r\!+\!1)},\ \ s(r)=\frac{r_0-r}{q} \label{solution} \\&&\!\!\!\!\!\!\!\!\!\!\!\!\!\!\!\!\!+\frac{2}{q\sqrt{3}}\arctan\left(\frac{2r-1}{\sqrt{3}}\right)
-\frac{2}{q\sqrt{3}}\arctan\left(\frac{2r_0-1}{\sqrt{3}}\right),\nonumber
\end{eqnarray}
where $r_0=r(s=0)$. The inversion of this formula, to find $r(s)$, gives a  transcendental equation. However, the asymptotic properties of the solution can be derived without solving the equation. When $s$ increases, $r$ decreases reaching $r=1$ at a finite value $s=s_*$ where,
\begin{eqnarray}&&\!\!\!\!\!\!\!\!\!\!\!\!\!\!\!\!
s_*=\frac{r_0-1}{q}+\frac{\pi}{3q\sqrt{3}}-\frac{2}{q\sqrt{3}}\arctan\left(\frac{2r_0-1}{\sqrt{3}}\right).
\end{eqnarray}
The situation of $r$ reaching $1$ would correspond to coalescence of the second and the third particles. This happens only asymptotically, as $s=s_*$ corresponds to infinite physical time, $t(s_*)=\infty$.
We have directly from Eq.~(\ref{solution}) that $s'(r=1)=0$ and $s''=(1-2r)/[q(r^2\!-\!r\!+\!1)^2]$ which gives,
\begin{eqnarray}&&\!\!\!\!\!\!\!\!\!\!\!\!\!\!\!\!
s(r)\approx s_*+(r-1)s'(r=1)+\frac{(r-1)^2s''(r=1)}{2}\nonumber\\&&\!\!\!\!\!\!\!\!\!\!\!\!\!\!\!\!\!
=s_*-\frac{(r-1)^2}{2q},\ \ (r-1)^2=2q(s_*-s). \label{rs}
\end{eqnarray}
We can find $x$ as a function of $r$ observing that,
\begin{eqnarray}&&\!\!\!\!\!\!\!\!\!\!\!\!\!\!\!\!
\frac{d\ln x}{dr}=\frac{d\ln x}{ds}\frac{ds}{dr}
=-\left(\frac{1}{r-1}+\frac{1}{r}\right)\frac{1}{r^2\!-\!r\!+\!1}.
\end{eqnarray}
Integration of the above gives (with $x_0=x(t=0)$),
\begin{eqnarray}&&\!\!\!\!\!\!\!\!\!\!\!\!\!\!\!\!
\ln \left(\frac{x}{x_0}\right)=\ln\frac{r_0(r_0-1)}{r_0^2\!-\!r_0\!+\!1}-\ln\frac{r(r-1)}{r^2\!-\!r\!+\!1},\label{rsol}
\end{eqnarray}
where we used,
\begin{eqnarray}&&\!\!\!\!\!\!\!\!\!\!\!\!\!\!\!\!
\int\left(\frac{1}{r-1}+\frac{1}{r}\right)\frac{1}{r^2\!-\!r\!+\!1}dr=\ln\frac{r(r-1)}{r^2\!-\!r\!+\!1}.
\end{eqnarray}
We find from Eq.~(\ref{rsol}) that,
\begin{eqnarray}&&\!\!\!\!\!\!\!\!\!\!\!\!\!\!\!\!
x=x_0\frac{r_0(r_0-1)(r^2\!-\!r\!+\!1)}{r(r-1)(r_0^2\!-\!r_0\!+\!1)}.
\end{eqnarray}
In the limit of large times where $r$ approaches $1$ from above we have,
\begin{eqnarray}&&\!\!\!\!\!\!\!\!\!\!\!\!\!\!\!\!
x\approx \frac{r_0(r_0-1)x_0}{(r-1)(r_0^2\!-\!r_0\!+\!1)}\approx \frac{r_0(r_0-1)x_0}{(r_0^2\!-\!r_0\!+\!1)\sqrt{2q(s_*-s)}}, \label{idst}
\end{eqnarray}
where we used Eq.~(\ref{rs}). Finally we restore the physical time using,
\begin{eqnarray}&&\!\!\!\!\!\!\!\!\!\!\!\!\!\!\!\!
\frac{dt}{ds}=x^3(s)\approx \frac{r_0^3(r_0-1)^3x_0^3}{(r_0^2\!-\!r_0\!+\!1)^3(2q)^{3/2}(s_*-s)^{3/2}},
\end{eqnarray}
and obtain,
\begin{eqnarray}&&\!\!\!\!\!\!\!\!\!\!\!\!\!\!\!\!
t(s)\approx \frac{r_0^3(r_0-1)^3x_0^3}{q(r_0^2\!-\!r_0\!+\!1)^3\sqrt{2q(s_*-s)}}.
\end{eqnarray}
We conclude from Eqs.~(\ref{rs}), (\ref{idst}) that the long-time asymptotic form of the solution is,
\begin{eqnarray}&&\!\!\!\!\!\!\!\!\!\!\!\!\!\!\!\!
x(t)=\frac{qt(r_0^2\!-\!r_0\!+\!1)^2}{r_0^2(r_0-1)^2x_0^2},\ \ r(t)=1+\frac{x_0^3r_0^3(r_0-1)^3}{qt(r_0^2\!-\!r_0\!+\!1)^3}.
\end{eqnarray}
This implies that the distance between the second and the third particles reaches a constant value at large times,
\begin{eqnarray}&&\!\!\!\!\!\!\!\!\!\!\!\!\!\!\!\!
x_{23}(t)=x(t)(r(t)-1)\approx \frac{x_0r_0(r_0-1)}{r_0^2\!-\!r_0\!+\!1}\nonumber\\&&\!\!\!\!\!\!\!\!\!\!\!\!\!\!\!\!\!=\frac{x_{13}(0)x_{23}(0)x_{12}(0)}{x_{13}(0)x_{23}(0)\!+\!x^2_{12}(0)}.\label{initial}
\end{eqnarray}
To conclude, we can write,
\begin{eqnarray}&&\!\!\!\!\!\!\!\!\!\!\!\!\!\!\!\!
x(t)=\frac{qt}{x_{23}^2},
\end{eqnarray}
in agreement with the form given by Eq.~(\ref{clust}). This relation proves the separation in cluster and faraway particle and provides the distance between the particles in the cluster as function of the initial conditions.

\begin{figure}[t!]
 \begin{center}
\includegraphics[width=1.\columnwidth]{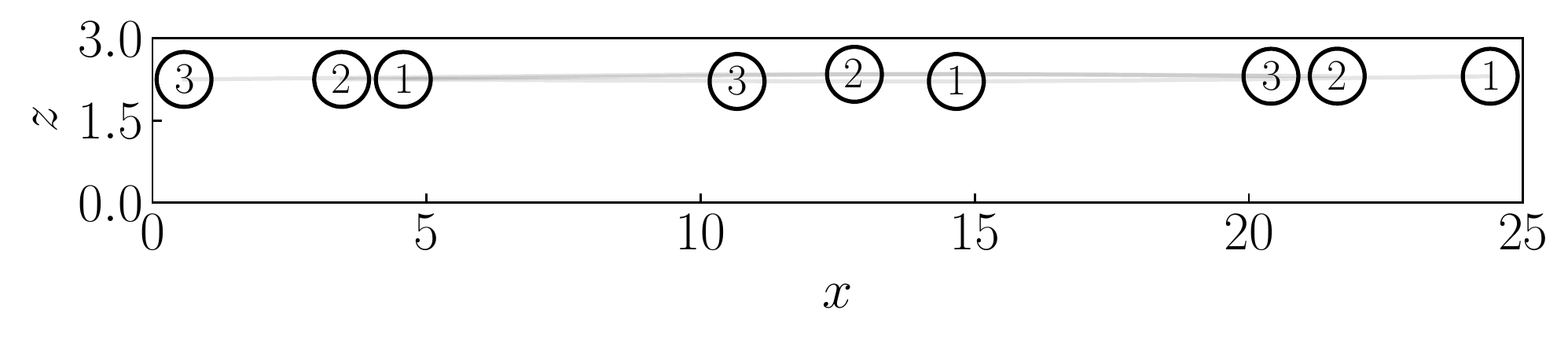}
\end{center}
  \caption{Pair exchange phenomenon as obtained from numerical simulations. Initially particles $1$ are $2$ are close and particle $3$ is trailing behind the pair. As a result of hydrodynamic interactions the trailing particle is catching up with the pair, while the leading particle breaks away from the newly formed pair, whereas  the trailing particles $2$ and $3$ are separated by the same distance as $1$ and $2$ were initially.}
 \label{fig: pair}
\end{figure}

The obtained formulas provide a theoretical explanation for the pair exchange observed in the experiments by \cite{tab0}, and confirmed by numerical
simulations as illustrated in Fig.\ \ref{fig: pair}. We consider initial conditions for which particles $1$ and $2$ are close and the third particle is trailing behind. In this case $x^2_{12}(0)\ll x_{13}(0),x_{23}(0)$ and Eq.~(\ref{initial}) becomes,
\begin{eqnarray}&&\!\!\!\!\!\!\!\!\!\!\!\!\!\!\!\!
x_{23}(t)=x_{12}(0).\label{initialp}
\end{eqnarray}
Thus, for long times the distance between the third and the second particle becomes equal to the initial distance between the first and the second particle, that is an exchange takes place.

We can also prove the separation in one cluster and one faraway particle for initial conditions where the particles are "almost" aligned, i.e.\  $y_{12}$ and $y_{13}$ are much smaller than the smallest of $x_{12}$ and $x_{13}$.
If the $y-$components of the particle positions are linearly ordered, the equations for $x_{ik}$ do not change and the evolution of $x_{ik}$ is as above. The $y-$components obey,
\begin{eqnarray}&&\!\!\!\!\!\!\!\!\!\!\!
\dot{y}_{12}\!=\!-\frac{2q y_{13}}{x_{13}^3}+\frac{2qy_{23}}{(x_{13}-x_{12})^3},\ \ \dot{y}_{23}=\frac{2q y_{13}}{x_{13}^3}-\frac{2q y_{12}}{x_{12}^3}\nonumber\\&&\!\!\!\!\!\!\!\!\!\!\!
\dot{y}_{13}\!=\!-\frac{2q y_{12}}{x_{12}^3}+\frac{2qy_{23}}{(x_{13}-x_{12})^3},
\end{eqnarray}
where $x_{ik}(t)$ are determined from the previous solution. Since $x_{12}$, $x_{13}$ grow linearly with time at large times then the asymptotic form of the solution is,
\begin{eqnarray}&&\!\!\!\!\!\!\!\!\!\!\!\!\!\!\!\!
y_{12}\!=\!y_{13}\!=\!\frac{2qc t}{x_0^3},\ \ y_{23}=c=const.\ \ \nonumber
\end{eqnarray}
This solution indicates that the first particle separates from the binary cluster at constant small but finite angle $\phi$ with respect to the $x-$direction given by,
\begin{eqnarray}&&\!\!\!\!\!\!\!\!\!\!\!\!\!\!\!\!
\phi=\frac{y_{12}}{x_{12}}\!\approx \frac{y_{13}}{x_{13}}\!\approx \!\frac{c}{x_0\sqrt{2}},
\end{eqnarray}
Self-consistency with the assumption of smallness of $y_{ik}$ demands that $c\ll x_0$. This assumption can be guaranteed by the smallness of the initial conditions on $y_{ik}$ because of the linearity of the equations in $y_{ik}$.

Thus we demonstrated the for initial conditions where the particles are aligned in the $x-$direction, or almost aligned, the solution at large times takes the form of a cluster of particles $2$ and $3$, those two initially upstream, with constant distance $\bm r_{23}$ and the first particle separating from the cluster according to Eq.~(\ref{clust}).

We formulate the hypothesis that any arbitrary initial configuration of three particles will lead at large times to a binary cluster and the third particle linearly separating from it. It seems that the evolution from any arbitrary initial conditions cannot be solved analytically, but only numerically: 
a reduction from four to three degrees of freedom can be obtained, but the resulting equations could not be solved.

To conclude, we describe the properties of cluster solutions assuming constant $\bm r_{23}$, of magnitude $r_{23}$, and angle $\phi$ with respect to the $x-$axis. Using Eq.~(\ref{clust}), we write
\begin{eqnarray}&&\!\!\!\!\!\!\!\!\!\!\!\!\!\!\!\!
\dot x_{12}\!\approx \dot x_{13}\approx \frac{q\cos(2\phi)}{r_{23}^2},\ \ \dot y_{12}\!\approx \dot y_{13}\approx \frac{q\sin(2\phi)}{r_{23}^2}. \nonumber
\end{eqnarray}
We can assume with no loss of generality that $x_2\geq x_3$ so that $\phi$ is in the range $-\pi/2\leq\phi\leq \pi/2$.
Four different solutions can therefore be identified, assuming the cluster at large time can be denoted as a point at the origin.
In the range $0\leq \phi\leq \pi/4$ the first particle leaves the cluster behind when going to infinity inside the first quadrant.
%
In the range $-\pi/4\leq \phi\leq 0$, the first particle leaves the cluster behind when going to infinity inside the second quadrant. In the case of $\pi/4\leq \phi\leq  \pi/2$ the cluster leaves the first particle behind and to the right. Finally in the case of $-\pi/2\leq \phi\leq  -\pi/4$ the cluster leaves the first particle behind and to the left.

It is clear from the above that the configuration with distant pairs or singlets of particles is stable: the singlets separate ballistically from the stable pairs which maintain the pair
distance constant.
In contrast, clusters of three and probably more particles are unstable.  Thus, we conjecture that under arbitrary initial conditions an arbitrary number of distant particles will
separate at large times into a collection of singlets and pairs if the solution is dilute. For suspensions of many particles this implies that hydrodynamic interactions increase the
probability of particles to be isolated or in pairs, rather than forming clusters composed of many particles. In dense suspensions, the interactions discussed above do not have
time to occur and we rather expect chaotic collisions of particles \cite{tl2014}.

\section{Concluding remarks}

In the present paper, we provided a boundary integral representation for the flow due to particles (rigid ones or droplets) freely-suspended in a channel flow. The particle number, size and shape and the inter-particle distances are arbitrary.
We have thus demonstrated the utility of this representation.

For an isolated particle, the proposed representation is useful for the study of the far-field flow. At the leading order, the far flow is dipolar flow with the dipole moment given by
a weighted integral of the stress tensor and the flow over the particle
surface. This defines the far flow completely. The flow was previously available only for strongly confined pancake-like droplets that almost block the channel in the vertical direction.  We also determine the dipole moment integral
numerically for neutrally buoyant rigid sphere. Further, we provide the multipole expansion from which the far flow can be found with any desired accuracy.

For close particles, the representation is helpful for the study of hydrodynamic interactions.
It demonstrates clearly that the range of validity of the lubrication theory is larger than expected from the usual
approach \cite{Batchelor,lubr,szeri,bruce}. Our representation  also solves the problem of matching the ideal flow holding far from the particles with the fully viscous flow near the particle surface. The solution is expressed in terms of
the unknown surface velocity and stress tensor. This sheds light on the use of the ideal flow approximation in previous works on disk-like particles \cite{tl2014,is} and helps to consider particles of other shapes.

We introduced the equation of motion of particles interacting at long distances, refining previous derivation for droplets \cite{tl2014} and extending it to the case of arbitrary,
possibly different, interacting particles. We solved the three-body problem of hydrodynamic interactions for the case of identical symmetric particles aligned in the stream-wise
direction. This solution provides theoretical support for the pair exchange phenomenon observed previously in experiments \cite{tab0}. We provide special solutions for the three-body problem and demonstrate that it is plausible that these solutions describe the long-time asymptotic evolution for arbitrary initial conditions. We further demonstrate the application of the theory to the many body problem.

We did not consider potential lateral migration induced by dydrodynamic interactions. It can be readily seen that a pair of distant spherical particles with different vertical coordinates will separate laterally because the coordinate-dependent mobility matrices and dipole moments will differ for these particles. The study of this instability will be the object of future work.

The representation proposed here is a good starting point for a mean-field description of strong hydrodynamic interactions of close particles. We notice after Eq.~(\ref{inrepmany}) that the effect of the interactions can be described by a  change of the stress-tensors and velocities at the particle surfaces. Thus the model description of the interaction boils down to the model description of surface stress tensors and velocity. This can be done by introducing the mean field $\nabla p$ whose direction can differ from the direction of the undisturbed flow. We assume that the stress tensor and flow on the surface of each particle is that for an isolated particle in the Poiseuille flow with pressure gradient $\nabla p$ (e.g., the particle velocity is equal to minus the mobility matrix times $\nabla p$). Using this in Eq.~(\ref{inrepmany}) one can find the flow in terms of $\nabla p$. A closed integral equation for $\nabla p$ can then be obtained from the Stokes flow equations. The study of this equation is planned as future work.

This study was majorly motivated by recent experiments on the formation of droplet clusters in a microfluidic channel \cite{tab0,tabeling}. The theoretical modeling in \cite{tabeling} assumed that flow-assisted clustering of weakly confined spherical droplets in close proximity is driven  by the combination of non-hydrodynamic (adhesive, e.g., depletion forces) and hydrodynamic interactions of dipolar nature similar to interactions of strongly confined (pancake-like) droplets in Hele-Shaw cells. The qualitative agreement between the results of the numerical simulations and experimental results in \cite{tabeling} suggested that \emph{ad hoc} modeling of hydrodynamic interactions by dipolar flow is admissible. The present study shows that far-field interactions of weakly confined droplets are indeed of dipolar nature, however their magnitude is too weak to lead to relative motion between freely suspended particles on the time scale of the experiment. Moreover, the present study, as well as the calculations of the interactions at small distances in \cite{tl2012,is}, suggest that hydrodynamic interactions  at close proximity cannot be described by dipolar flows. We thus believe that the reason for the qualitative agreement between the numerical results and the experiment (using unknown magnitude of the adhesive forces as an adjustable parameter) is that the adhesive force diverges at contact dominating the particle dynamics. The dipolar hydrodynamic interactions provided the source of sliding (tangential) motions of the particles necessary for the particle rearrangement and not provided by the adhesive (radial) forces. However, the particular functional form (e.g., dipolar or other) of these interactions seems to be of minor importance as long as these provide some tangential mobility. An accurate \emph{quantitative} predictive theory of flow-assisted clustering requires the knowledge of the near-field hydrodynamic interactions, including an accurate treatment of the non-uniform flow near the inlet. This will be the object of a future work.

\section*{Acknowledgement}
The work is supported by the Microflusa project. The Microflusa project receives funding from the European Union Horizon 2020 research and innovation programme
under Grant Agreement No. 664823.

{}
\begin{appendices} 

\section{Calculation of the far-field term $f_l(\bm x_0)$}\label{f}

In this appendix, we present a direct calculation of the far-field term $f_l(\bm x_0)$ in Eq.~(\ref{in}). We find from volume integration of Eq.~(\ref{fd}) the boundary term,
\begin{eqnarray}&&\!\!\!
f_l(\bm x_0)=\frac{1}{8\pi\eta} \int_{-L}^L dy\int_0^h dz \left[S_{il}(L-x_0, y-y_0, z, z_0)\right.\nonumber\\&&\!\!\!\left.\sigma_{ix}(L, y, z)-S_{il}(-L-x_0, y-y_0, z, z_0)\sigma_{ix}(-L, y, z)\right]\nonumber\\&&\!\!\!+\frac{1}{8\pi\eta} \int_{-L}^L dx\int_0^h dz \left[S_{il}(x-x_0, L-y_0, z, z_0)\right.\nonumber\\&&\!\!\!\left.\sigma_{iy}(x, L, z)-S_{il}(x-x_0, -L-y_0, z, z_0)\sigma_{iy}(x, -L, z)\right]\nonumber\\&&\!\!\!-\frac{1}{8\pi} \int_{-L}^L dy\int_0^h dz \left[T_{ilx}(L-x_0, y-y_0, z, z_0)u_{i}(L, y, z)\right.\nonumber\\&&\!\!\!\left.-T_{ilx}(-L-x_0, y-y_0, z, z_0)u_{i}(-L, y, z)\right]-\frac{1}{8\pi}\nonumber\\&&\!\!\! \times\int_{-L}^L dx\int_0^h dz \left[T_{ily}(x-x_0, L-y_0, z, z_0)u_{i}(x, L, z)\right.\nonumber\\&&\!\!\!\left.-T_{ily}(x-x_0, -L-y_0, z, z_0)u_{i}(x, -L, z)\right];\ \ L\to\infty.\nonumber
\end{eqnarray}
The Stokeslet decays exponentially in the $z-$direction so $f_z=0$. To find the remaining components we use
\begin{eqnarray}&&\!\!\!\!\!\!\!\!\!\!
\sigma_{ix}(L, y, z)=-L\nabla_x p^0\delta_{ix}+\frac{(2z-h)\nabla_x p^0\delta_{iz} }{2}.
\end{eqnarray}
We find that $f_x$ is determined by the asymptotic solution for channel flow and is not affected by the presence of the spherical particle,
\begin{eqnarray}&&
f_x(\bm x_0)=-\frac{L\nabla_x p^0}{8\pi\eta} \int_{-L}^L dy\int_0^h dz \left[S_{xx}(L, y-y_0, z, z_0)\right.\nonumber\\&&\left.+S_{xx}(-L, y-y_0, z, z_0)\right]\nonumber\\&&-\frac{\nabla_x p^0}{8\pi\eta} \int_{-L}^L dx\int_0^h dz x\left[S_{yx}(x-x_0, L, z, z_0)\right.\nonumber\\&&\left.-S_{yx}(x-x_0, -L, z, z_0)\right]\nonumber\\&&-\frac{1}{8\pi} \int_{-L}^L dy\int_0^h dz \frac{z(z-h)\nabla_x p^0}{2\eta}\left[T_{xxx}(L, y-y_0, z, z_0)\right.\nonumber\\&&\left.-T_{xxx}(-L, y-y_0, z-z_0)\right]\nonumber\\&&-\frac{1}{8\pi} \int_{-L}^L dx\int_0^h dz \frac{z(z-h)\nabla_x p^0}{2\eta}\left[T_{xxy}(x-x_0, L, z, z_0)\right.\nonumber\\&&\left.-T_{xxy}(x-x_0, -L, z-z_0)\right];\ \ L\to\infty.
\end{eqnarray}
Thus, this must be the unperturbed channel flow as readily verified. Rescaling the integration variable by $L$ and keeping leading order terms we have,
\begin{eqnarray}&&\!\!\!\!\!\!\!\!\!\!
f_x(\bm x_0)=-\frac{L^2\nabla_x p^0}{8\pi\eta} \int_{-1}^1 dy\int_0^h dz \left[S_{xx}(L, Ly, z, z_0)\right.\nonumber\\&&\!\!\!\!\!\!\!\!\!\!\left.+S_{xx}(-L, Ly, z, z_0)\right]\nonumber\\&&\!\!\!\!\!\!\!\!\!\!-\frac{L^2\nabla_x p^0}{8\pi\eta} \int_{-1}^1 dx\int_0^h dz x\left[S_{yx}(Lx, L, z, z_0)\right.\nonumber\\&&\!\!\!\!\!\!\!\!\!\!\left.-S_{yx}(Lx, -L, z, z_0)\right]\nonumber\\&&\!\!\!\!\!\!\!\!\!\!-\frac{L}{8\pi} \int_{-1}^1 dy\int_0^h dz \frac{z(z-h)\nabla_x p}{2\eta}\left[T_{xxx}(L, Ly, z, z_0)\right.\nonumber\\&&\!\!\!\!\!\!\!\!\!\!\left.-T_{xxx}(-L, Ly, z-z_0)\right]\nonumber\\&&\!\!\!\!\!\!\!\!\!\!-\frac{L}{8\pi} \int_{-1}^1 dx\int_0^h dz \frac{z(z-h)\nabla_x p}{2\eta}\left[T_{xxy}(Lx, L, z, z_0)\right.\nonumber\\&&\!\!\!\!\!\!\!\!\!\!\left.-T_{xxy}(Lx, -L, z-z_0)\right].
\end{eqnarray}
So far the calculation involved the complete Stokeslet solution. To determine $f_l$ we can use the asymptotic form of the Stokeslet at large distances, which is  for the stress tensor,
\begin{eqnarray}&&\!\!\!\!\!\!\!\!\!\!
T_{ilk}=-24\frac{r_l}{\rho^2}(z_0/ h^2)(1-z_0/h)\delta_{ik}+O\left(\frac{1}{\rho^2}\right).
\end{eqnarray}
Using these formulas for $S_{ik}$ and $T_{ilk}$, one can write
\begin{eqnarray}&&\!\!\!\!\!\!\!\!\!\!
f_x=\frac{\nabla_x p^0}{8\pi\eta} \int_{-1}^1 dy\int_0^h dz \frac{24z(z-h)z_0(z_0-h)(y^2-1)}{(1+y^2)^2 h^3}\nonumber\\&&\!\!\!\!\!\!\!\!\!\!-\frac{\nabla_x p}{4\pi\eta} \int_{-1}^1 dx\int_0^h dz x^2\left[
\frac{24z(z-h)z_0(z_0-h)}{(1+x^2)^2 h^3}\right]
\nonumber\\&&\!\!\!\!\!\!\!\!\!\!-\frac{1}{4\pi} \int_{-1}^1 dy\int_0^h dz \frac{z(z-h)\nabla_x p}{2\eta}\left[24\frac{z_0(z_0-h)}{(1+y^2)h^3}\right]\nonumber.
\end{eqnarray}
Integrating over $z$, this can be written as,
\begin{eqnarray}&&\!\!\!\!\!\!\!\!\!\!
f_x=\frac{\nabla_x p z_0(z_0-h)}{2\eta}\left( \int_{-1}^1 \frac{dy}{\pi(1+y^2)}\left[\frac{2}{1+y^2}-1\right]\right.\nonumber\\&&\!\!\!\!\!\!\!\!\!\!\left.+\frac{1}{\pi} \int_{-1}^1 dx \frac{2x^2}{(1+x^2)^2}+\frac{1}{\pi} \int_{-1}^1 dy \left[\frac{1}{(1+y^2)}\right]\right).
\end{eqnarray}
Performing the integrals we confirm that indeed $f_x$ is the flow given by Eq.~(\ref{ps}).

\section{Numerical integration of $\bm s$}\label{numm}

We compute the weighted dipole moment $\bm s$ in Eq.~(\ref{dlpa}) by directly simulating a rigid spherical particle of radius $a$ transported in a doubly-periodic channel using the immersed boundary method (IBM), see \cite{ibm,ruth,picano} for more details and validations.

In the IBM, there are two meshes; one Eulerian mesh for the flow and one
Lagrangian mesh for the moving particle. The two meshes are coupled through a multidirect forcing scheme that ensures the approximate no-slip/no-penetration condition on the particle surface.

The motion of the particle is described by the Newton-Euler equations, given for the translational velocity  by
Eq.\ \eqref{eqmp}).
The equation does not contain the gravitational force which is assumed to be balanced either by the particle interactions with the bottom wall (the case of particle near the wall) or by buoyancy (the case of density-matched particle). We also assume that particle-wall collisions are absent (\textit{cf.} Eq.\ (3a) in \cite{ibm}).

The flow outside the particle is governed by the incompressible Navier-Stokes (NS) equations with the no-slip boundary conditions on the surface of the rigid particle described in connection with Eq.\ \eqref{ds}. Although the steady state flow obeys the Stokes equations, computation of the transients demands inclusion of the time derivative in the NS equations. In our simulations, the full NS equations are computed at a small Reynolds number ($\sim10^{-1}$), \textit{viz.}
\begin{equation}
    Re\bigg(\frac{\partial {\bm u}}{\partial t} + \nabla \cdot ({\bm u \bm u}) \bigg)=
    \nabla \cdot {\bm \sigma} + {\bm f},
    \label{NS}
\end{equation}
where $Re=\rho_f U_b (2a)/\eta$ is the Reynolds number, $U_b$ the channel bulk velocity, $\rho_f$ ($=\rho_p$) the fluid density and ${\bm f}$ the IBM force enforcing that the no-slip boundary condition (in this formulation pressure is rescaled by $Re$.) We discretize these equations using a second-order finite volume scheme. Finding the flow at given translational and rotational particle velocities we obtain the viscous stress which is used for updating these velocities as in Eq.\ \eqref{eqmp}. Numerically, the LHS of Eq.\ \eqref{eqmp} is computed at each time step by summing the forces exerted on all the Lagrangian points, in addition to the volumetric forces inside the particle (see Eq.\ (8a) in \cite{ibm} for the full expression). In our case, this is simply
\begin{equation}
     m\frac{d\bm v}{dt}  \approx
    -\sum_{l=1}^{N_l} {\bm F}_l \Delta V_l + \rho_f\frac{d}{dt}\bigg(\int_{V_p} {\bm u}dV \bigg),
    \label{ibm}
\end{equation}
where 
$-{\bm F}_l$ the force acting on the $l$ Lagrangian point  centred at a shell element of volume $\Delta V_l$, and $N_l$ the total number of Lagrangian points.

At the steady state,
\begin{equation}
    \int_{S}
   {\bm \sigma} \cdot \bm {dS}\approx -\sum_{l=1}^{N_l} {\bm F}_l \Delta V_l,
    \label{ibm 1}
\end{equation}
corresponding to the solution of the
steady state the Stokes equations, Eq.\ \eqref{ds}.

Provided that the interpolation and spreading between ${\bm f}_{i,j,k}$ and $-{\bm F}_l$ preserves the local stress, we obtain the dimensionless weighted dipole moment $\tilde{\bm s}$ needed to compute particle interacions as
\begin{equation}
    \tilde{{\bm s}} = -\sum_{l=1}^{N_l} z_l(h-z_l) {\bm F_{l}} \Delta V_l.
    \label{formula}
\end{equation}
The dimensional ${\bm s}$ is thus $\eta U_b (2a)^3\tilde{\bm s}$.\\

\end{appendices}

\end{document}